\documentclass[a4paper,fleqn,usenatbib]{mnras}
\usepackage{graphicx}
\usepackage{amsmath}
\usepackage{amssymb}
\usepackage{txfonts}
\usepackage{subcaption}
\usepackage{upgreek}
\usepackage{rotating}
\usepackage{tabularx}
\usepackage[usenames,dvipsnames,svgnames,table]{xcolor}
\usepackage{txfonts}
\DeclareSymbolFont{cmletters}{OML}{cmm}{m}{it}
\DeclareMathSymbol{v}{\mathalpha}{cmletters}{"76}
\newcommand{\GG}[1]{}

\newcommand{\RedeclareMathOperator}[2]{\renewcommand{#1}{}\let#1\relax\DeclareMathOperator{#1}{#2}}

\newcommand{\hammer}{{\texttt{H-AMR}}}

\newcommand{\bhoss}{{\texttt{BHOSS}}}

\newcommand\simless\lesssim
\newcommand\simgreat\gtrsim

\newcommand{\weak}{{\texttt{Weak}}}
\newcommand{\strong}{{\texttt{Strong}}}
\newcommand{\weakC}{{\texttt{Weak}-$\epsilon_{\rm C}$}}
\newcommand{\weakPIC}{{\texttt{Weak}-$\epsilon_{\rm PIC}$}}
\newcommand{\strongC}{{\texttt{Strong}-$\epsilon_{\rm C}$}}
\newcommand{\strongPIC}{{\texttt{Strong}-$\epsilon_{\rm PIC}$}}
\newcommand{\chandra}{{\textit{Chandra}}}
\newcommand{\nustar}{{\textit{NuSTAR}}}
\newcommand{\sgra}{{Sgr~A$^*$}}

\title[Non-thermal flaring in Sagittarius A*]{General relativistic MHD simulations of non-thermal flaring in Sagittarius A*}
\author[Chatterjee et al.]{K. Chatterjee$^{1,2}$\thanks{E-mail: koushik.chatterjee@cfa.harvard.edu}, S. Markoff$^{1,3}$, J. Neilsen$^{4}$, Z. Younsi$^{5,6}$, G. Witzel$^{7}$, A. Tchekhovskoy$^{8}$, \newauthor D. Yoon$^{1}$,  A. Ingram$^{9}$, M. van der Klis$^{1}$, H. Boyce$^{10}$, T. Do$^{11}$, D. Haggard$^{10}$ \newauthor \& M. A. Nowak$^{12}$\\
$^{1}$Anton Pannekoek Institute for Astronomy, University of Amsterdam, Science Park 904, 1098 XH Amsterdam, The Netherlands\\
$^{2}$Black Hole Initiative, Harvard University, 20 Garden Street, Cambridge, MA 02138, USA \\
$^{3}$Gravitation Astroparticle Physics Amsterdam (GRAPPA) Institute, University of Amsterdam, Science Park 904, 1098 XH Amsterdam, The Netherlands\\
$^{4}$Department of Physics, Villanova University, 800 Lancaster Avenue, Villanova, PA 19085, USA\\
$^{5}$Mullard Space Science Laboratory, University College London, Holmbury St.~Mary, Dorking, Surrey, RH5 6NT, United Kingdom\\
$^{6}$Institut f{\"u}r Theoretische Physik, Goethe-Universit{\"a}t Frankfurt, Max-von-Laue-Stra{\ss}e 1, D-60438 Frankfurt am Main, Germany\\
$^{7}$Max Planck Institute for Radio Astronomy, Auf dem Hügel 69, D-53121 Bonn (Endenich), Germany\\
$^{8}$Center for Interdisciplinary Exploration \& Research in Astrophysics (CIERA), Physics \& Astronomy, Northwestern University, Evanston, IL 60202, USA\\
$^{9}$Department of Physics, Astrophysics, University of Oxford, Denys Wilkinson Building, Keble Road, Oxford, OX1 3RH, UK\\
$^{10}$McGill Space Institute and Department of Physics, McGill University, 3600 rue University, Montreal, QC H3A 2T8, Canada\\
$^{11}$UCLA Galactic Center Group, Physics and Astronomy Department, University of California, Los Angeles, CA 90024, USA\\
$^{12}$Department of Physics, Washington University, CB 1058, One Brookings Drive, St. Louis, MO 63130-4899, USA
}
\date{Accepted XXX. Received YYY; in original form ZZZ}
\pubyear{2019}
\begin{document}
\label{firstpage}
\pagerange{\pageref{firstpage}--\pageref{lastpage}} 
\maketitle

\begin{abstract}

\sgra{} exhibits regular variability in its multiwavelength emission, including daily X-ray flares and roughly continuous near-infrared (NIR) flickering. The origin of this variability is still ambiguous since both inverse Compton and synchrotron emission are possible radiative mechanisms. The underlying particle distributions are also not well constrained, particularly the non-thermal contribution. In this work, we employ the GPU-accelerated general relativistic magnetohydrodynamics (GRMHD) code \hammer{} to perform a study of flare flux distributions, including the effect of particle acceleration for the first time in high-resolution 3D simulations of \sgra{}. For the particle acceleration, we use the general relativistic ray-tracing (GRRT) code \bhoss{} to perform the radiative transfer, assuming a hybrid thermal$+$non-thermal electron energy distribution. We extract $\sim 60$~hr lightcurves in the sub-millimetre, NIR and X-ray wavebands, and compare the power spectra and the cumulative flux distributions of the lightcurves to statistical descriptions for \sgra{} flares. Our results indicate that non-thermal populations of electrons arising from turbulence-driven reconnection in weakly magnetised accretion flows lead to moderate NIR and X-ray flares and reasonably describe the X-ray flux distribution while fulfilling multiwavelength flux constraints. These models exhibit high rms\% amplitudes, $\gtrsim 150\%$ both in the NIR and the X-rays, with changes in the accretion rate driving the 230~GHz flux variability, in agreement with \sgra{} observations.

\end{abstract}


\section{Introduction}
\label{sec:introduction}

The extreme physical conditions in the vicinity of accreting black holes (BHs) present a unique opportunity to study the acceleration of particles in conditions unattainable on Earth. Magnetised plasma turbulence, instabilities, and shocks occurring naturally in accretion flows and outflows are all potential processes that trigger particle acceleration. Sagittarius A$^*$ (\sgra{}), the supermassive BH (SMBH) candidate at the centre of our galaxy, due to its proximity, presents an excellent opportunity to test current theories about particle acceleration near BHs against high-quality observational data. 
Intensive monitoring of \sgra{} has led to accurate measurements of stellar orbits yielding a BH mass of $M_{\rm BH}=4.1 \times 10^6\,M_{\odot}$ \citep{Ghez:05,Gillessen:17, Gravity:18_S2} at a distance of $D_{\rm BH}=8.15$~kpc \citep[e.g.,][]{Ghez:08,Boehle:16,Reid:19} from the Earth, and resulted in a systematic study of its emission in the radio, millimetre (mm), near-infrared (NIR) and X-ray wavebands \citep[e.g., see][and references therein]{Genzel:2010}. \sgra{} is a remarkably faint SMBH (luminosity $L_{\rm bol}\sim 10^{-9}L_{\rm Edd}$, where $L_{\rm Edd}$ is the Eddington luminosity) that accretes gas at an estimated rate of $\dot{M}\sim 10^{-9}-10^{-7}\,M_{\odot}$~yr$^{-1}$ \citep{Bower:03,Marrone:07,Wang:2013}. Almost $99$\% of the accreted gas at the Bondi scale is lost due to turbulence and/or outflows by the time the flow reaches the black hole \citep{Wang:2013}. At such a low accretion rate the accretion flow can be expected to be radiatively inefficient (e.g., \citealt{Yuan_2003}; also see \citealt{Yuan_2014} and references therein). In spite of its low luminosity, \sgra{} is one of our best opportunities to study a SMBH via its interaction with accreting material, and it plays a crucial role in our understanding of extreme gravitational environments. 

Observations at multiple wavelengths over the previous two decades have constrained the quiescent spectrum of \sgra{}. In the sub-millimetre (sub-mm) band, \citet{Bower_2019} found a spectral index of $\alpha_{\nu}\simeq -0.31$ (where the flux density is $F_{\nu}\propto \nu^{\alpha_{\nu}}$) with the peak flux lying between 1 and 2~terahertz. Using a thermal synchrotron emission model to account for both the sub-mm and the NIR flux, \citet{Bower_2019} estimates an electron temperature of $T_{\rm e} \simeq 10^{11}$~K along with a small magnetic field strength of $\sim 10-50$~G in the inner accretion flow, consistent with previous semi-analytic results \citep[e.g.,][]{Falcke:2000,Markoff_2001_sgra,Yuan_2002}. In the quiescent state, the X-rays seem to be dominated by thermal bremsstrahlung from the Bondi-scale accretion flow \citep[e.g.,][]{quataert:2002,Baganoff:03,Yuan_2003}.

Apart from its low-luminosity quiescent state, \sgra{} regularly displays fluctuations in flux across multiple frequencies, most prominently in the NIR and X-ray bands, which are often correlated with each other \citep[e.g.,][]{Eckart:2004,Dodds-Eden_2009,Boyce_2019}. Ever since \citet{Baganoff:01} reported the first detection of an X-ray flaring event in \sgra{} with the \textit{Chandra X-ray Observatory}, the SMBH has been the target of multiple observational campaigns (e.g., the 3 Ms 2012 \chandra{} X-ray Visionary Project\footnote{\href{http://www.sgra-star.com}{http://www.sgra-star.com}} and the \textit{Neil Gehrels Swift Observatory}\footnote{\href{https://swift.gsfc.nasa.gov}{https://swift.gsfc.nasa.gov}} monitoring campaign; \citealt{Degenaar:2015}). In 2019, \citet{Do_2019} and \citet{Haggard_2019} reported the largest flares yet detected from \sgra{} at 2.12~$\mu$m (NIR) with the Keck Telescope and in the 2-8~keV energy band (X-ray) with \chandra{}, respectively. Further, \nustar{} observations confirmed that \sgra{}'s X-ray flares have higher energy extensions \citep[e.g.,][]{Barriere_2014} with a luminosity of $L_{3-79~\rm keV}\sim (0.7-4.0)\times 10^{35}$~erg s$^{-1}$ and photon index $\Gamma=2.2\pm 0.1$ \citep{Zhang_2017} similar to the 2-8~keV \chandra{} \citet{Nowak_2012} photon index ($\Gamma\simeq 2.0$). The X-ray emission of \sgra{} is the lowest among observed low-luminosity SMBHs, allowing the detection of low-flux stochastic flaring events in the innermost regions of the accretion flow. It is still an open question of whether this is due to the absence of a pronounced jet feature that usually dominates the X-ray emission in other accreting SMBHs. The immense amount of observational data allows us to investigate these pivotal questions about the plasma conditions in \sgra{}. 

Most semi-analytical studies are agnostic about the exact mechanism behind particle acceleration in \sgra{} \citep[e.g.,][]{Quataert:1999:ADAF_spectra,Ozel_2000, Markoff_2001_sgra, Liu_2001, Yuan_2003, Markoff_2005_sgra, Li2015, Connors_2017}, whether it be from shocks or magnetic reconnection. Non-thermal synchrotron emission is generally favoured over synchrotron-self-Compton (SSC) to be the source of simultaneous NIR and X-ray flaring events since only a few local parameters, such as the acceleration efficiency or the power-law index, need to change in order to transition from the quiescent phase to a flare state \citep[e.g.,][]{Dibi_2014}. Additionally, when simultaneous NIR/X-ray flares are observed the required steepening of the slope between these bands is a challenge to fit with SSC models but fits expectations for a cooling break in synchrotron radiation \citep[e.g.,][]{Markoff_2001_sgra, Dodds-Eden_2009, Dodds-Eden_2010, Dibi_2014, Ponti_2017}. For SSC flare models, \citet[][]{Dibi_2014} showed that flares can only occur if the accretion rate and/or the global magnetic field strength changes. However, the low radio and sub-mm flux variability may rule out large changes in the accretion rate during observed flaring states. With more simultaneous NIR/X-ray surveys, such as \citet[][]{Boyce_2019}, flare flux distributions provide the potential to statistically differentiate between synchrotron and SSC models.

Flare flux probability distribution functions (PDFs) are a useful tool to understand the possible physical mechanisms behind NIR versus X-ray flaring, without requiring individual fits to the properties of individual flares such as flare duration, shape and time delays. The NIR cumulative distribution function (CDF) consists of a log-normal component that captures the low-flux distribution \citep[][]{Witzel_2012, Witzel_2018} with an additional power-law for the high fluxes \citep[][]{Dodds-Eden_2010}. In this paper, we take the CDF from \citet[][]{Do_2019}, where in addition to the log-normal component from \citet{Witzel_2018}, there is a power-law tail with index $\sim 2$. However, we should note here that the power-law index is still debated as \citet{Dodds-Eden_2010} found an index of $\sim 1.7$ while \citet[][]{Witzel_2012} and \citet{Witzel_2018} found a steep index of $\sim 3.2-3.6$ \footnote{Some of these papers actually quote the slope of the flux PDFs instead of the CDFs. For a power-law, the slope is reduced by 1 when transforming from the PDF to the CDF.}. The X-ray CDF can be modelled as a Poissonian with a power-law tail \citep[][]{Neilsen_2015}. The power-law index for the NIR CDF is $1.92$, which is close to the \citet{Do_2019} NIR CDF slope. Taking the \citet{Witzel_2012} NIR CDF slope of 3.2 and the \citet[][]{Neilsen_2015} X-ray CDF slope of 1.92, \citet{Dibi_2016} shows that the large difference between the slopes of NIR and X-ray cumulative distribution functions is not well described by simpler synchrotron scenarios and leaves the door open for a fuller exploration of models involving both SSC from a thermal electron population and synchrotron from a non-thermal electron population. Hence, further studies of the relation of the NIR and X-ray CDF slope is required to disentangle common processes that could give rise to both flare populations.

The origin of large NIR flares such as the one reported in \citet{Do_2019} is still unknown. In \citet{Do_2019}, the authors suggest that an increase in the accretion rate could explain the exceptionally high flux. Unfortunately, simultaneous observations at sub-mm are not available to confirm this theory. As an alternative to a change in the accretion rate, \citet{Gutierrez_2020} demonstrates that large NIR flares could have a non-thermal origin in the form of a magnetised blob of plasma, i.e., a plasmoid. Local particle-in-cell simulations show that plasmoids can naturally form as a result of relativistic magnetic reconnection in environments that are prevalent in accretion discs as well as the jet boundary \citep[e.g.,][]{sironi2015, Ball_2018_PIC, Hakobyan2019}, but whether such plasmoids grow to the sizes required to explain the enormous flux is not yet known. The computational demands of these simulations prohibit the exploration of bulk flow effects on the microscopic plasmoid behaviour, requiring alternate schemes to incorporate the global turbulence of the accretion flows. 

The idea that plasmoids are responsible for NIR flaring gained even more traction with the \citet{Gravity:18_hotspot} detection of three NIR flares consistent with hotspots orbiting at a distance of $\sim 6-10$ gravitational radii ($r_{\rm g}\equiv GM_{\rm BH}/c^2$, where $G$ and $c$ are the gravitational constant and the speed of light respectively) from the BH with an inclination of approximately $140^{\circ}$ degrees and an orbital period $\sim 115$~minutes \citep{Gravity:20:orbit}. This detection points towards localised mechanisms such as magnetic reconnection and electron heating behind the NIR flaring, indicating the need to understand the small-scale activity of the accretion flow via numerical simulations \citep[e.g., see][]{Dexter:2020:NIR_MAD,Porth:2020:NIR_MAD}.

Over the past decades, the theoretical astrophysics community has increasingly used general relativistic magnetohydrodynamic (GRMHD) simulations of accreting BHs to produce a more self-consistent description of accretion dynamics. GRMHD simulations usually employ the single fluid approximation, assuming that the ion temperature dominates the flow temperature and that the ions are in a thermal distribution. With the additional simplifying assumption of neutral hydrogen, we assume each proton is accompanied by an electron. Therefore, we require a model that describes the electron temperature in postprocessing \citep[e.g.,][]{Howes:2010,Rowan:2017,Werner_2018}. Along with GRMHD quantities, the choice of the electron temperature model is important to generate multiwavelength spectra using general relativistic ray-tracing (GRRT) codes, and thus, enable us to compare simulated data to the observed spectrum of \sgra{} \citep[e.g.,][]{Dexter_2009,Moscibrodzka_2009, Dexter_2010, Dibi_2012, Shcherbakov_2012, Drappeau_2013, moscibrodzka_2013, Chan_2015, Mao:2017, Davelaar2018, Anantua_2020}. A computationally more expensive alternative approach is to evolve the electron thermodynamics along with the gas evolution self-consistently within a GRMHD simulation \citep[e.g.,][]{ressler_2015, Ryan2017, Chael_2018, Dexter_2020,Mizuno_2021}. 

Using the robust framework of GRMHD$+$GRRT methods \citep[e.g.,][]{Porth:19,Gold_2020}, quite a few studies have tried to investigate the properties of the accretion flow that lead to flares. The 2D GRMHD simulations in \citet{Drappeau_2013} consider thermal SSC origins of X-ray flaring. \citet{Chan_2015} was the first to use 3D GRMHD simulations that invoke thermal bremsstrahlung emission to model undetected X-ray flares \citep[$\simless 10\%$ of the quiescent X-ray flux;][]{Neilsen_2013}, thought to come from the inner accretion flow. Further, \citet{Ball_2016} and \citet{Mao:2017} used a hybrid thermal+non-thermal synchrotron model \citep{Ozel_2000} while \citet{Davelaar2018} employed a $\kappa$-distribution model to study \sgra{}'s X-ray emission. However, the 2D restriction and/or the relatively low resolution of these simulations might result in spurious numerical artefacts. Indeed, \citet{Dexter:2020:NIR_MAD} and \citet{Porth:2020:NIR_MAD} employed high resolution 3D GRMHD simulations to argue that magnetically saturated BHs release small scale magnetic eruptions leading to the formation of orbiting NIR features, a possible mechanism to explain GRAVITY-observed flares. \citet[][]{Porth:2020:NIR_MAD} found that the observed hotspots orbit faster than the simulated NIR features, hinting at super-Keplerian motions \citep[e.g.,][]{Matsumoto:2020}. Although these GRMHD studies restrict their scope to reconnection features in the disc, possible NIR flare models do not yet preclude outflowing features. Outflowing magnetic reconnection zones at wind or jet boundaries, when viewed in projection \citep[e.g.,][]{Nathanail_2020,Ripperda_2020,Ball_2021}, could appear to move as the observed hotspots, despite the apparent absence of a well-collimated outflow in \sgra{} \citep[though the size constraints of the central sub-mm source might allow for a jet; e.g., see][]{Markoff_2007, Issaoun_2019}. The highly anticipated results of the Event Horizon Telescope \citep{doeleman2008,EHT_paperI} observations of \sgra{} promise to reveal much more about the horizon-scale structure and add to the plethora of observational data. Despite the recent advances made in modelling flares, no 3D high resolution GRMHD simulation has yet tackled the inconsistency of \sgra{}'s NIR/Xray flare flux distribution slopes raised in \citet{Dibi_2016} and whether turbulence-driven non-thermal activity can explain the overall behaviour of NIR and X-ray flares.

To address these questions, we employ the state-of-the-art GPU-accelerated GRMHD code \hammer{} \citep{liska_hamr2020_arxiv} to simulate BH accretion discs and jets, together with the GRRT code \bhoss{} \citep{younsi_2019_polarizedbhoss} to generate synchrotron-only spectra assuming a hybrid thermal+non-thermal electron distribution, and do not account for inverse Compton scattering. For the first time, we evolve fully 3D simulations at high resolutions to calculate \sgra{} lightcurves over a significantly long time period to study the role of disc and jet turbulence in determining the observed flux variability. The long lightcurves thus produced also provide a key perspective beyond spectral properties. To that end we extract NIR/X-ray flare statistics from the simulations, providing a view of GRMHD BH discs complementary to other recent theoretical papers that focus on magnetised features to explain the NIR flaring \citep[e.g.,][]{Dexter:2020:NIR_MAD,Gutierrez_2020,Petersen:2020,Porth:2020:NIR_MAD,Ball_2021}. For the NIR/X-ray flare statistics, we primarily focus on comparing cumulative flux distributions both in the NIR \citep{Do_2019} and the X-rays \citep{Neilsen_2015}, restricting our analysis to the inner accretion flow (within $50~r_{\rm g}$) that has marginally reached inflow equilibrium \citep[using the criteria given in][]{Narayan:2012}. We provide a detailed overview of our numerical methods in Sec.~\ref{sec:grmhd_setup} and \ref{sec:GRRT_setup}, present and discuss our results in Sec.~\ref{sec:results} and \ref{sec:variability}, and subsequently, conclude in Sec.~\ref{sec:conclusions}.

\begin{table}
\begin{centering}
\renewcommand{\arraystretch}{1.3}
\begin{tabularx}{\columnwidth}{l}
\hline\hline
\textbf{GRMHD parameters} \\
\hline\hline
\end{tabularx}
\begin{tabularx}{\columnwidth}{l c c c c c}
Model & a      & Resolution                                             & $r_{\rm in}$  & $r_{\rm max}$  & $r_{\rm out}$ \\
      &        &($N_{\rm r}\times N_{\rm \theta}\times N_{\rm \varphi}$)& [$r_{\rm g}$] & [$r_{\rm g}$]  & [$r_{\rm g}$] \\
\hline
All   & 0.9375 & $240\times144\times256$                                & 12.5          & 25             & $1000$        \\
\hline
\hline
\end{tabularx}
\begin{tabularx}{\columnwidth}{l c c c c }
Model     &  B-flux & $\phi_{\rm BH}$ & $Q$-factor                         & $t_{\rm sim}$       \\
          & strength &               & $(Q_{r}, Q_{\theta}, Q_{\varphi})$ &[$10^4 r_{\rm g}]/c$ \\
\hline
\weak{}   & Weak-field    & 3.8 & (7.1, 8.1, 29.5)                   & 2.93                \\
\strong{} & Strong-field  & 39.7 & 34.5, 26.1, 85.4)                 & 3.34                \\
\hline\hline
\end{tabularx}
\begin{tabularx}{\columnwidth}{l}
\textbf{GRRT  parameters} \\
\hline\hline
\end{tabularx}
\begin{tabularx}{\columnwidth}{l c c c c}
Model & FOV                           & Image           & \sgra{} $M_{\rm BH}$ \& $D_{\rm BH}$ & Inclination \\
      & [$r_{\rm g}\times r_{\rm g}$] & resolution      & [$M_{\odot}$, kpc]       & [degrees]   \\
\hline
All   & $50\times 50$               & $1024\times 1024$ & $4.1\times 10^6, 8.15$   & $85^{\circ}$, $25^{\circ}$ \\
\hline
\hline
\end{tabularx}
\begin{tabularx}{\columnwidth}{l c c c c c}
Model    & $R_{\rm high}$ & $R_{\rm low}$ & Accretion rate          & $\epsilon_{\rm C}$ & $p_{\rm C}$   \\
         &                &               & [$M_{\odot}$~yr$^{-1}$] &                   \\
\hline
\weak{}  & 10             & 10            & $3.69\times 10^{-8}$    & $0.01$             & $2$\\
\strong{}& 40             & 10            & $2.96\times 10^{-8}$    & $2.5\times 10^{-4}$& $2$\\
\hline\hline
\end{tabularx}

\caption{Top row: GRMHD parameters common to simulation models used in this work - dimensionless BH spin (a), simulation grid resolution, disc inner radius ($r_{\rm in}$), disc pressure-maximum radius ($r_{\rm max}$), and outer grid radius ($r_{\rm out}$).
2\textsuperscript{nd} row: simulation model names, disc magnetic flux strength, time-averaged dimensionless magnetic flux through the event horizon (see Sec.~\ref{sec:grmhd_results} for definition), time-averaged density-weighted volume-averaged MRI quality factors ($Q_{r,\theta,\varphi}$, see Sec.~\ref{sec:grmhd_setup} for definition), and total simulation time ($t_{\rm sim}$) in $r_{\rm g}/c$.
3\textsuperscript{rd} row: GRRT parameters common to radiative models - observer field of view (FOV), GRRT image resolution in pixels, \sgra{} BH mass and distance used for GRRT calculations, and source inclination angle with respect to observer. 
Bottom row: radiative model names, $R_{\rm high}$ and $R_{\rm low}$ parameters for the electron temperature prescription ( Eqn.~\eqref{eq:R_parameter}), the time-averaged accretion rate, the electron acceleration efficiency coefficient and the power-law index for the constant power-law injection scheme (see Sec.~\ref{sec:plasmoid_model_const}). We calculate the time-average quantities over a period of $\sim 60~$hrs.}

\label{tab:models}
\end{centering}
\end{table}

\section{Simulation setup}
\label{sec:grmhd_setup}
\hammer{} \citep{liska_tilt_2018,chatterjee2019,Porth:19,liska_hamr2020_arxiv} evolves the GRMHD equations set in a fixed Kerr spacetime, specifically in logarithmic Kerr-Schild (KS) coordinates. \hammer{} makes use of advanced techniques such as adaptive mesh refinement (AMR), static mesh de-refinement, local adaptive time-stepping and a staggered mesh setup for evolving the magnetic field (see \citealt{liska_hamr2020_arxiv} for more details about these methods, and \citealt{Porth:19} for comparisons to results from other current GRMHD codes for a standard BH accretion disc problem). We adopt the geometrical unit convention, taking $G=c=1$, and normalise our length scale to the gravitational radius $r_{\rm g}$. The GRMHD simulation grid is axisymmetric, logarithmically-spaced in $r$, and uniform in $\theta$ and $\phi$, and extends from $r=1.21r_{\rm g}$ to $10^3r_{\rm g}$. We use static mesh de-refinement to reduce the number of cells in $\phi$ around the polar axis by a factor of 2 \citep[similar to][but without stretching the polar cells in $\theta$]{liska_tilt_2018}. We did not make use of AMR in this work. We take the BH spin parameter to be $a=0.9375$ and therefore, our inner radial boundary is inside the event horizon radius ($r_{\rm H}=1.347r_{\rm g}$). The grid resolution is $N_{r}\times N_{\theta}\times N_{\varphi}\equiv 240\times 144\times 256$. The resolution in $\theta$ sufficiently resolves moderately thick discs \citep[scale height $h/r\sim 0.1-0.3$ near the BH;][]{Porth:19} with 5-13 cells. We use outflowing radial boundary conditions (BCs), transmissive polar BCs \citep[see supplementary information in][]{liska_tilt_2018} and periodic $\varphi$-BCs. 

The accretion disc is set up in the form of the standard \cite{fis76} hydrostatic torus rotating around the spinning BH (refer to Table~\ref{tab:models} for torus and other model specifications). A non-relativistic ideal gas equation of state is assumed: the gas pressure $p_{\rm gas}=(\gamma_{\rm ad}-1)u_{\rm g}$, where $\gamma_{\rm ad}=5/3$ and $u_{\rm g}$ is the internal energy. We perform two simulations, one that leads to a weak jet (model \weak{}) and the other a relatively strong jet (model \strong{}). We assume a single poloidal loop in the initial disc magnetic field configuration for both simulations, indicated by the magnetic vector potential ($\Vec{A}$):

\begin{eqnarray}
	\weak: A_{\phi} &\propto& \left\{ 
	  \begin{array}{ll}
		  \rho - 0.2,   & {\rm if}~~ \rho > 0.2, \\
		  0,            & {\rm otherwise}.  
	  \end{array} \right. \\
	\strong{}: A_{\phi} &\propto& \left\{ 
	  \begin{array}{ll}
		  (\rho - 0.05)^2r^5,   & {\rm if}~~ \rho > 0.05, \\
		  0,            & {\rm otherwise},  
	  \end{array} \right.
\end{eqnarray}

\noindent where $\rho$ is the rest-mass gas density in code units. The magnetic field strength in the initial condition is normalised by setting $\max(p_{\rm g})/\max(p_{\rm B})=100$, where $p_{\rm B}=b^2/2$ is the magnetic pressure in Heaviside-Lorentz units and $b$ is the co-moving magnetic field strength. Solving the GRMHD equations provides the gas density, internal energy, velocities and magnetic field components per grid cell. Assuming a hydrogen-only, electron-proton plasma, protons dominate the gas density and internal energy, and therefore, we need to assume an electron distribution function and temperature in postprocessing.

The grid resolution is the same for both simulations and is sufficient to resolve the magnetorotational instability \citep[MRI;][]{bal91} in the disc. We use the standard MRI quality factors $Q_{r, \theta, \varphi}=<2\pi v_{\rm A}^{r, \theta, \phi}>_{\rho}/<\Delta^{r,\theta,\phi} \Omega>_{\rho}$ to measure the number of cells resolving the largest MRI wavelength, volume-averaged over the disc (using the gas density $\rho$ as the weight in the average; \citealt{Chatterjee_2020}). In the definition of $Q$, $v_{\rm A}^i$, $\Delta^i$ and $\Omega$ are the Alfv\'en speed in the $i$-th direction, the corresponding cell size and the fluid angular velocity, respectively. While we adequately resolve the MRI in model \strong{} with $Q$ values above 25 \citep[a minimum of 10 is usually quoted for convergence of disc parameters, see][]{hgk11, Porth:19}, model \weak{} is only marginally resolved due to weaker magnetic fields in the disc. A brief summary of the simulation details is given in Table~\ref{tab:models}. 

Current grid codes are prone to numerical errors when solving the GRMHD equations for gas density and internal energy within the vacuous jet funnel. These errors are due to gas either being expelled as an outflow or accreted via the BH's gravity, leaving behind a vacuum region with extremely high magnetisations that GRMHD codes fail to deal with, hence requiring the use of an ad-hoc density floor model. We set a minimum gas density limit of $\rho_{\rm min} c^2 \geq \max\left[p_{\rm B}/50,\,2\times10^{-6}c^2(r/r_{\rm g})^{-2}\right]$ and a minimum internal energy limit of $u_{\rm g, min} \geq \max\left[p_{\rm B}/150,\,10^{-7}c^2(r/r_{\rm g})^{-2\gamma_{\rm ad}}\right]$, mass-loading the jet funnel according to the implementation described in \citet{ressler_2017}. Gas thermodynamics is unreliable in the jet funnel since small errors in $B$ propagate as large errors in internal energy due to the huge scale separation between magnetic and thermal energies. Therefore, we do not account for any emission coming the jet, which we define as regions with $\sigma_{\rm M}=b^2/\rho c^2>1$, where $\sigma_{\rm M}$ is the magnetisation.

\section{Radiative transfer model}
\label{sec:GRRT_setup}

In this section, we describe our model for the electron distribution function, using the general relativistic ray-tracing (GRRT) code \bhoss{} \citep{younsi_2012,younsi_2016, younsi_2019_polarizedbhoss} to calculate the corresponding synchrotron emission. We generate multiwavelength images and spectra of both simulations (scaled to the mass and distance of \sgra{}) at a cadence of $5~r_{\rm g}/c$, accounting for all emission within $50~r_{\rm g}$ only. We employ a hybrid thermal$+$non-thermal electron distribution, assuming  two different acceleration models for the non-thermal synchrotron emission. In order to speed up our GRRT calculations, we make use of fitting functions for the synchrotron emissivity and absorption of a relativistic thermal Maxwell-J\"uttner distribution from \citet{Leung_2011} and a non-thermal power-law distribution from \citet[][]{Fouka_2014} (see Appendix~\ref{sec:Fouka} for more details).

\subsection{Thermal synchrotron modelling}

First, we calculate the electron temperature $T_{\rm e}$ via the \citet[][]{Moscibrodzka_2016} prescription based on turbulent heating models that gives us the ion-electron temperature ratio ($R\equiv T_{\rm i}/T_{\rm e}$) in the form,

\begin{equation}
    R=\frac{R_{\rm low}+R_{\rm high}\beta_{\rm P}^2}{1+\beta_{\rm P}^2},
    \label{eq:R_parameter}
\end{equation}{}

\noindent where $\beta_{\rm P}$ is the plasma-$\beta$, defined as the ratio of the gas and magnetic pressures. Plasma-$\beta$ varies both in time and space with typically large $\beta$ values in the disc and small $\beta$ in the jet. Assuming that the contribution of the electrons to the total gas pressure is negligible, we can calculate the electron temperature as:
\begin{equation}
    T_{\rm e}= \frac{m_{\rm p}p_{\rm gas}}{\rho k_{\rm B} R}, 
    \label{eq:elec_temp}
\end{equation}{}
\noindent where $m_{\rm p}$ and $k_{\rm B}$ are the proton mass and the Boltzmann constant respectively. We adopt $R_{\rm high}$ and $R_{\rm low}$ values for each simulation model such that we are able produce an average spectrum which resembles \sgra{}'s sub-mm to NIR spectrum. The specific values of $R_{\rm high}$ and $R_{\rm low}$ are given in Table~\ref{tab:models}. Using the electron temperature prescription, we can calculate the thermal synchrotron spectrum from our GRMHD models. 

The total thermal energy density is given by, 
\begin{equation}
    U_{\rm th} \, = \, n_{\rm th} \, u_{\rm th} \, = \, n_{\rm th} \, f(\Theta_{\rm e}) \, \Theta_{\rm e} \, m_{\rm e} \, c^2,
    \label{eqn:thermal_energy_density}
\end{equation}
\noindent where $n_{\rm th}=\int^{\infty}_{1} (dN_{\rm th}/d\gamma) \, d\gamma$ is the total thermal electron number density. Here, $\Theta_{\rm e} \, (\equiv k_{\rm B}T_{\rm e}/m_{\rm e}c^2)$ is the dimensionless electron temperature with $m_{\rm e}$ as the electron mass. \citet{Gammie_1998} gives $f(\Theta_{\rm e})$ in a simplified form,  
\begin{equation}
    f(\Theta_{\rm e})=\frac{6+15\Theta_{\rm e}}{4+5\Theta_{\rm e}}.
\end{equation}
\noindent The expression for $f(\Theta_{\rm e})$ gives us the thermal energy density $u_{\rm th}=(3/2)\Theta_{\rm e}$ for small $\Theta_{\rm e}$, i.e., non-relativistic temperatures, and $u_{\rm th}=3\Theta_{\rm e}$ for large relativistic temperatures. 

\subsection{Non-thermal synchrotron modelling} 

We consider a simple non-thermal synchrotron model where electrons are accelerated via magnetic dissipation, inspired by the early hybrid thermal$+$non-thermal model of \citet{Ozel_2000}. Such a treatment for the electron distribution function is similar to several other prior works, e.g., \citet{Markoff_2001_sgra, Broderick_2010,dexter_2012,Mao:2017,Connors_2017}. In practice, we take a portion of the available thermal electron population and create a non-thermal population with a given total energy density. The co-moving energy density in the non-thermal population ($U_{\rm nth}$) is given by,

\begin{equation}
    U_{\rm nth}=\int^{\gamma_{\rm max}}_{\gamma_{\rm min}} \gamma\frac{dN_{\rm nth}}{d\gamma}m_{\rm e} c^2 d\gamma,
    \label{eqn:nth_energy_density}
\end{equation}{}

\noindent where we have a power-law distribution $dN_{\rm nth}/d\gamma\propto\gamma^{-p}$. Here, $\gamma$ is the electron Lorentz factor, $p$ is the power-law distribution index, and $\gamma_{\rm max}$ and $\gamma_{\rm min}$ are the maximum and minimum electron Lorentz factors in the distribution respectively. We tie $\gamma_{\rm min}$ to the peak of the Maxwellian distribution\footnote{We note that other authors have used the condition $dN_{\rm th}/d\gamma(\gamma_{\rm min})=dN_{\rm nth}/d\gamma(\gamma_{\rm min})$ to smoothly connect the thermal and non-thermal distributions and calculate $\gamma_{\rm min}$ \citep[e.g.,][]{Ozel_2000,Yuan_2003, Mao:2017}.}:

\begin{equation}
    \gamma_{\rm min}=\gamma_{\rm pk}\simeq 1+u_{\rm th}/m_{\rm e}c^2, 
\end{equation}

\noindent which provides a physical normalisation for the power-law since PIC simulations demonstrate that electrons accelerate out of the thermal pool. We set $\gamma_{\rm max}$ to be
\begin{equation}
    \gamma_{\rm max}=\eta_{\gamma}\gamma_{\rm min}.
\end{equation}

\noindent \citet{Markoff_2001_sgra} found that $\gamma_{\rm max}\gtrsim 10^5$ is required to explain the \citet{Baganoff:01} X-ray flare, thus motivating an assumed $\eta_{\gamma}$ of $10^4$ to achieve large X-ray fluxes in regions with high electron temperatures. 

Next, instead of a single power-law, we include synchrotron cooling (refer to Appendix Eqn.~\eqref{eqn:brokenPL}) by calculating the break Lorentz factor $\gamma_{\rm br}$. To get $\gamma_{\rm br}$ for a particular cell, we equate the local synchrotron cooling timescale $t_{\rm sync}$ to the advection timescale $r/|v^r|$ as follows: 
\begin{equation}
    t_{\rm sync}\equiv\frac{6\pi m_{\rm e}c^3}{\sigma_{\rm T}b^2 \gamma_{\rm br}v_{\rm br}^2}=\frac{r}{|v^r|}.
\end{equation}
\label{eqn:gamma_br}
\noindent Here, $v^r$ and $\sigma_{\rm T}$ are the bulk fluid radial velocity and the Thomson cross-section, while the electron velocity $v_{\rm br}/c\equiv \sqrt{1-1/\gamma_{\rm br}^2}\approx 1$. Synchrotron cooling is crucial for our analysis since previous semi-analytical work such as \citet[][]{Dodds-Eden_2009} has shown that a broken power-law is necessary explain simultaneous NIR/X-ray flares with synchrotron.

In highly magnetised regions, we accelerate electrons out of the thermal pool to a power-law distribution using two different injection models and calculate the radiative output of each GRMHD model (see Table~\ref{tab:models}).

\subsubsection{Constant power-law injection: $\epsilon_{\rm C}$ model}
\label{sec:plasmoid_model_const}

Our first acceleration model assumes that the total non-thermal electron energy density is a fraction of the available magnetic field energy \citep[e.g.,][]{Broderick_2010, dexter_2012}. We assume a constant injected power-law index $p$ of 
\begin{equation}
    p=p_{\rm C}=2
\end{equation}
\noindent motivated by both observations \citep[e.g.,][]{Nowak_2012} and semi-analytical modelling \citep[e.g.,][]{Connors_2017} of \sgra{}. Since we expect efficient particle acceleration to occur in magnetically dominated regions, we use a criterion for the acceleration efficiency that promotes non-thermal activity in regions where the magnetic energy dominates over the rest-mass energy, i.e., where $p_{\rm B}\gtrsim \rho c^2$ \citep{Broderick_2010}. Keeping our criterion in mind, we assume a total non-thermal energy density:

\begin{equation}
    U_{\rm nth}=\left(\frac{2\epsilon_{\rm C}}{1+\exp(\rho c^2/p_{\rm B})}\right)p_{\rm B}=\epsilon_{\rm C, eff}p_{\rm B}.
    \label{eqn:broderick_nth}
\end{equation}{}

\noindent The above equation, combined with Eqn.~\eqref{eqn:nth_energy_density}, gives us the non-thermal electron number density for each grid cell, and captures changes in the magnetic energy density that might lead to NIR and X-ray variability, given a constant efficiency coefficient $\epsilon_{\rm C}$. As \citet{Broderick_2010} state, Eqn.~\eqref{eqn:broderick_nth} reduces to $U_{\rm nth}=\epsilon_{\rm C}p_{\rm B}$ in highly magnetised zones. We note that the reduced form of Eqn.~\eqref{eqn:broderick_nth} is the same as that employed in \citet{dexter_2012}, where the authors considered the jet launching region in M87. Particle-in-cell (PIC) simulations provide a self-consistent explanation of the generation of non-thermal activity by resolving the formation and evolution of plasmoids within current sheets using a fully kinetic framework \citep[e.g.,][]{Sironi_2014, Guo_2014, Werner_2018, Ball_2018_PIC}. In the next section, we describe a PIC-motivated radiation model.
\begin{figure}  
\begin{centering} 
\vspace{-10pt}
    \includegraphics[width=\columnwidth,trim=0cm 0cm 0cm 0cm,clip]{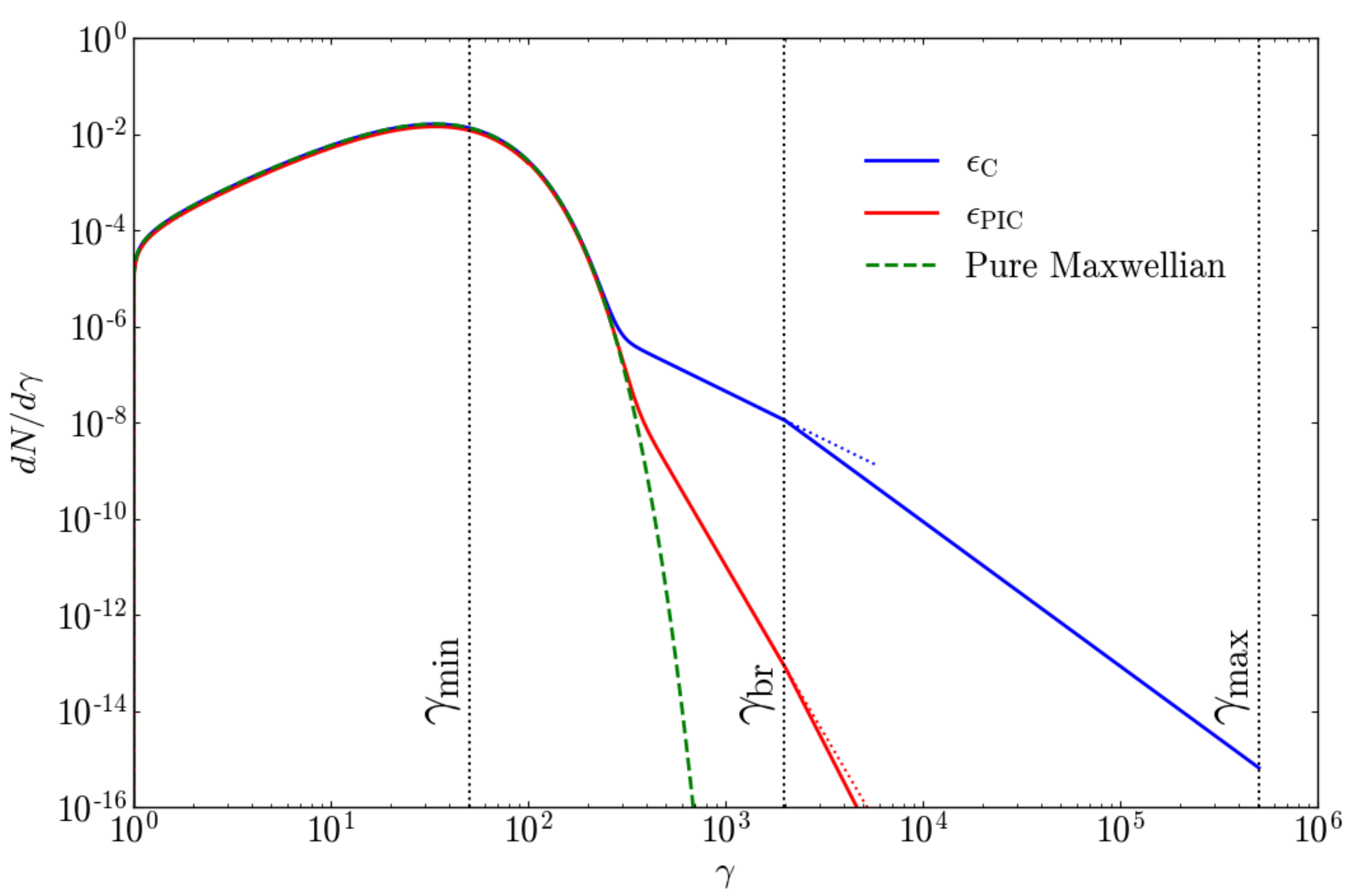}
    \caption{The electron distribution function for a cell in the magnetised inner accretion flow for both types of non-thermal models considered: constant power-law $\epsilon_{\rm C}$ and varying power-law $\epsilon_{\rm PIC}$. We assume an electron temperature $T_{\rm e}=10^{11}$~K, magnetisation $\sigma_{\rm M}=0.5$, $R_{\rm low}=10$,  $R_{\rm high}=40$, magnetic field $B=40$~G, here located at a radius of $5~r_{\rm g}$ from the BH. For the power-law distributions, we show the minimum and maximum electron Lorentz factors and the break Lorentz factor. The dotted blue and red lines illustrates the break in the power-law due to synchrotron cooling. For comparison, we also show the relativistic Maxwell-J\"uttner distribution for these parameters.}
    \label{fig:edf}
\end{centering} 
\end{figure}

\begin{figure*}  
\begin{centering}
    \includegraphics[width=\textwidth,trim=0cm 0cm 0cm 0cm,clip]{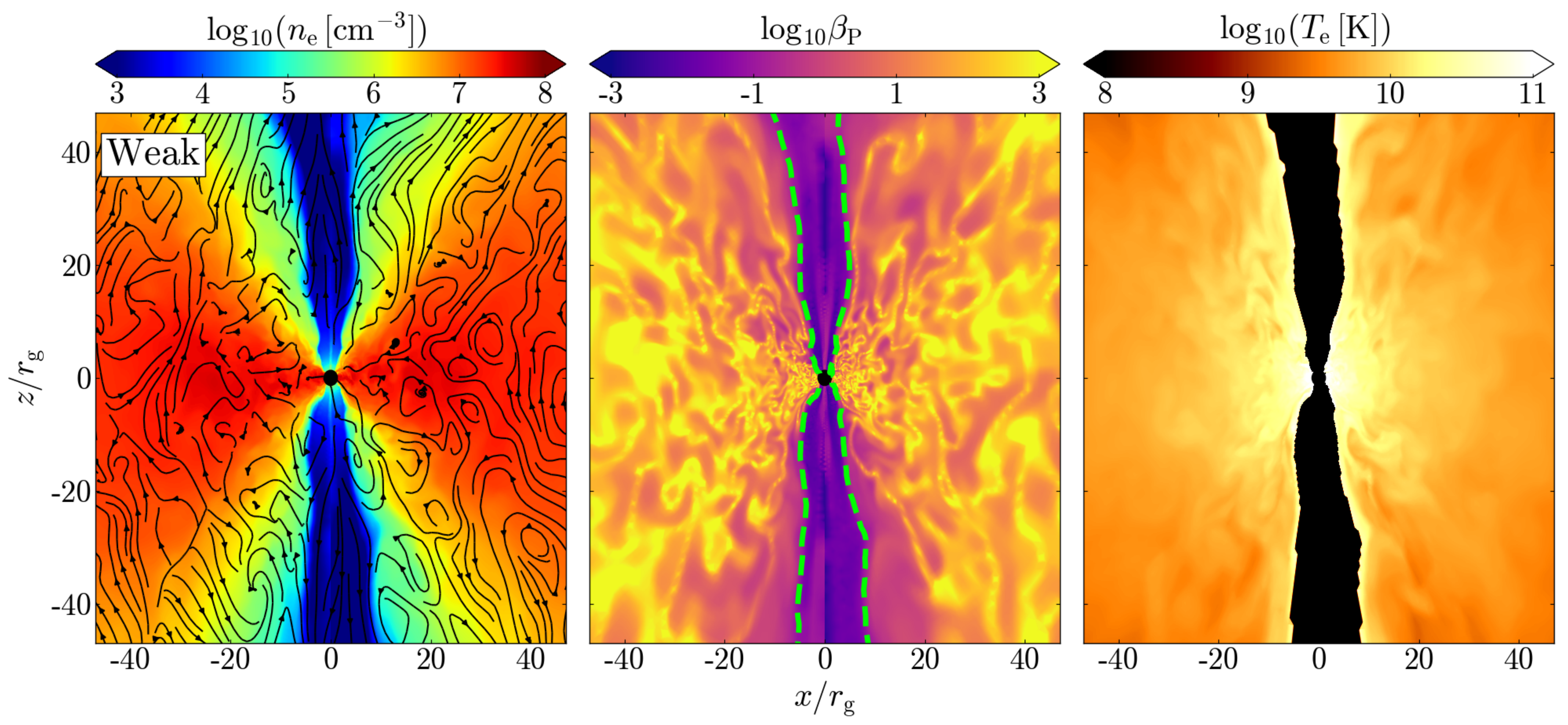}
    \includegraphics[width=\textwidth,trim=0cm 0cm 0cm 0cm,clip]{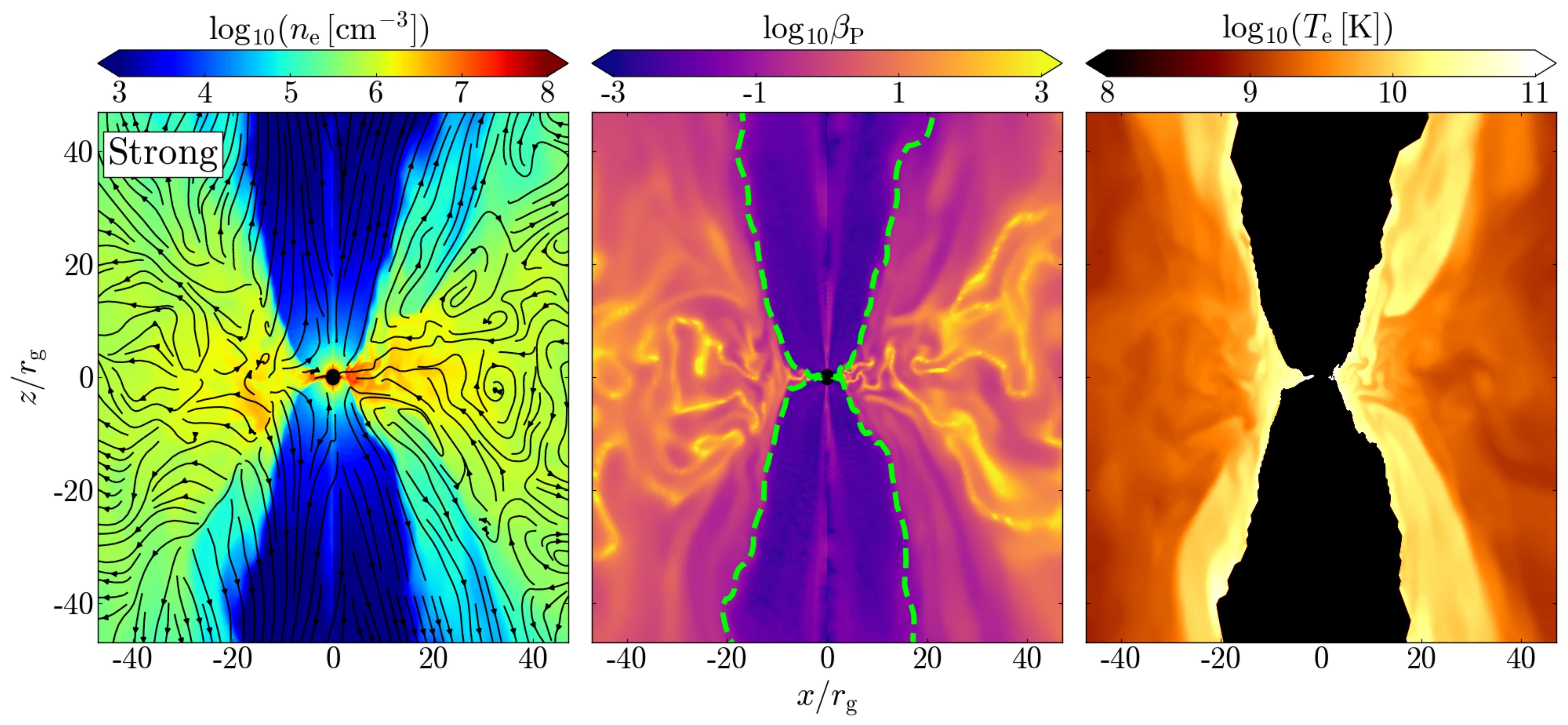}
    \caption{Strongly magnetised discs lead to wider jets, magnetically dominated winds and lower disc densities and temperatures as compared to their weak disc counterparts. We show x-z cross-sections of the electron number density $n_{\rm e}$ (with velocity streamlines in black), plasma-$\beta$ and electron temperature $T_{\rm e}$ from our GRMHD simulations: (top row) weak-field disc model \weak{} and (bottom row) strong-field disc model \strong{}, scaled to \sgra{} according to Table~\ref{tab:models}. We also indicate the jet boundary (approximated as magnetisation $\sigma_{\rm M}=1$) in green for the plasma-$\beta$ plots. Further, the jet is cut out (black region) in the $T_{\rm e}$ plots, since we do not account for the jet spine emission in our radiative scheme. Snapshots are taken at approximately 28,000 $r_{\rm g}/c$.}
    \label{fig:grmhd}

\end{centering} 
\end{figure*}

\subsubsection{Varying power-law injection: $\epsilon_{\rm PIC}$ model}
\label{sec:plasmoid_model_ball}

In the previous section, we relied on a physically motivated ad-hoc prescription for the acceleration efficiency and a constant power-law index, remaining agnostic about the accelerating process. In the case of magnetic reconnection, PIC simulations suggest that both of these quantities are dependent on the surrounding conditions, such as plasma-$\beta_{\rm P}$ and the magnetisation $\sigma_{\rm M}$ ($\equiv b^2/\rho c^2$) \citep[e.g.,][and references therein]{Werner_2018, Ball_2018_PIC}, especially in the trans-relativistic regime (i.e., $\sigma_{\rm M}\sim 1$). Current 3D GRMHD simulations lack the resolution required to resolve the small-scale structure of plasmoids, or magnetised blobs of gas, that form as a result of magnetic reconnection in current sheets. Only recently have we seen plasmoid evolution in high resolution 2D GRMHD simulations \citep{Nathanail_2020, Ripperda_2020}. While it is conceivable that current sheets may be resolvable using advanced simulation grids such as adaptive meshes, for this study, we rely on current sheets and particle acceleration prescriptions from PIC parameter surveys to generate the variability seen in the X-ray emission of \sgra{}.

\begin{figure*}  
\begin{centering}
\vspace{-10pt}
    \includegraphics[width=\textwidth,trim=0cm 0cm 0cm 0cm,clip]{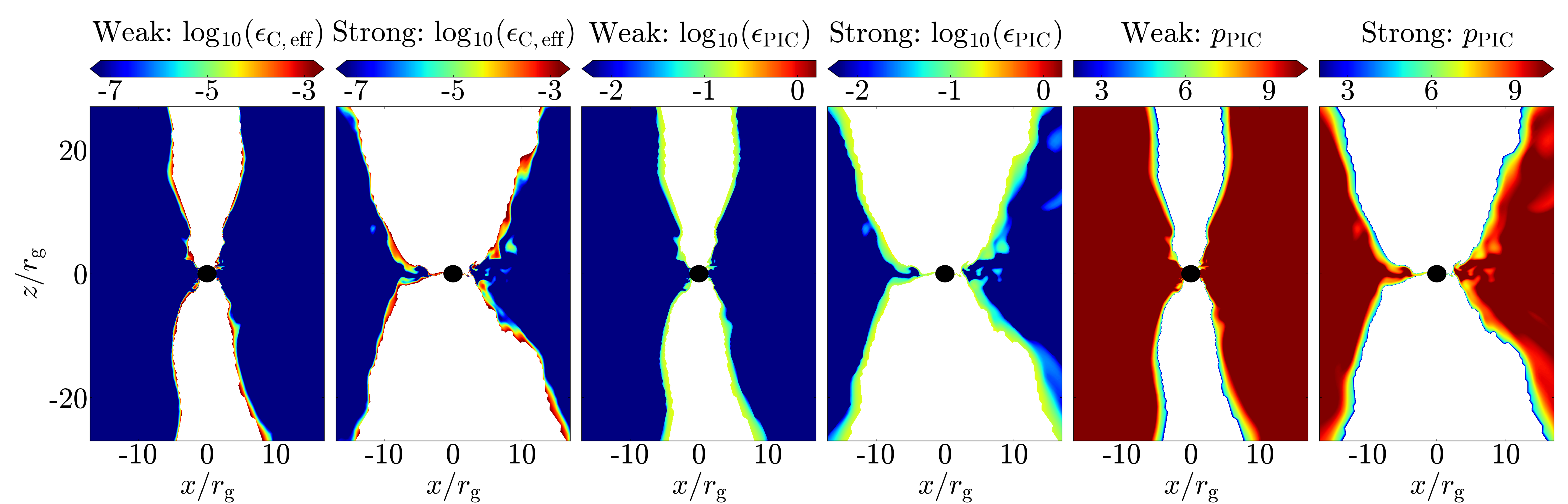}
    \caption{A 2D cross-section of the acceleration efficiencies for each radiative model: constant power-law injection $\epsilon_{\rm C}$ and varying power-law injection $\epsilon_{\rm PIC}$ schemes for both \weak{} and \strong{} GRMHD models. We also show the the power-law index $p_{\rm PIC}$ for the varying power-law injection model. Here, $\epsilon_{\rm C, eff}$ is the effective efficiency from Eqn.~\eqref{eqn:broderick_nth} and $\epsilon_{\rm PIC}$ is from Eqn.~\eqref{eqn:efficiency}. Most of the non-thermal activity occurs in the jet sheath or within a few $r_{\rm g}$ of the BH where efficiencies are high enough and the power-law index becomes harder. We do not account for any emission from the jet funnel (the white region in each plot).}
    \label{fig:eff_pl}
\end{centering} 
\end{figure*}

We incorporate the non-thermal electron acceleration prescriptions for magnetic reconnection given by \citet{Ball_2018_PIC}. \citet{Davelaar2019} used the same acceleration prescriptions to incorporate non-thermal particles in their simulations of M87. The primary difference in our approach is that whereas \citet{Davelaar2019} employed the use of the relativistic $\kappa-$distribution function \citep{Xiao:2006}, applying only the power-law index prescription of \citet{Ball_2018_PIC}, we calculate the thermal and non-thermal components of the synchrotron emission separately and account for both PIC-motivated power-law indices and efficiencies. Such an approach removes two degrees of freedom from our first radiative scheme. 

\citet{Ball_2018_PIC} tracks the evolution of the non-thermal electron distribution in a reconnecting layer embedded in an ambient plasma-$\beta_{\rm P}$ and magnetisation $\sigma_{\rm M}$ (hereafter referred to as $\beta_{\rm P, amb}$ and $\sigma_{\rm M, amb}$). For our simulations, we calculate the ambient values for each cell in the entire simulation grid by taking the average over the nearest 2 cells in each direction as follows:
\begin{align}
    \beta_{\rm P, amb}^{-1}&=\sum_{(i-2,j-2,k-2)}^{(i+2,j+2,k+2)}\beta_{\rm P}^{-1}(r,\theta,\phi)\\
    \sigma_{\rm M, amb}&=\sum_{(i-2,j-2,k-2)}^{(i+2,j+2,k+2)}\sigma_{\rm M}(r,\theta,\phi).
\end{align}{}
\noindent In our simulations, current sheets would appear in the disc/sheath as regions with small values of $\beta_{\rm P}^{-1}$ (i.e., regions where magnetic fields reconnect and the field strength drops) encapsulated by regions of large $\beta_{\rm P}^{-1}$ \citep[e.g.,][]{Ripperda_2020}. A grid cell with large values of $\beta_{\rm P, amb}^{-1}$ and $\sigma_{\rm M, amb}$ would presumably contain a current sheet within it and should exhibit a power-law tail in the electron distribution.

\citet{Ball_2018_PIC} gives the power-law electron distribution slope $p$ and the non-thermal acceleration efficiency $\epsilon_{\rm PIC}$ in terms of $\beta_{\rm P, amb}$ and $\sigma_{\rm M, amb}$:
\begin{equation}
    p=A_p+B_p \tanh{C_p\beta_{\rm P, amb}},
\label{eqn:power-law}
\end{equation}
\noindent where $A_p=1.8+0.7/\sqrt{\sigma_{\rm M, amb}}$, $B_p=3.7\sigma_{\rm M, amb}^{-0.19}$, $C_p=23.4\sigma_{\rm M, amb}^{0.26}$, and,
\begin{equation}
    \epsilon_{\rm PIC} =\frac{\int^{\infty}_{\gamma_{\rm pk}}(\gamma - 1)\left[\frac{dN}{d\gamma}-\frac{dN_{\rm th}(\gamma,\Theta_{\rm e})}{d\gamma}\right]d\gamma}{\int^{\infty}_{\gamma_{\rm pk}}(\gamma - 1)\frac{dN}{d\gamma}d\gamma}.
    \label{eqn:efficiency_ball}
\end{equation}{}  
\noindent Since the total electron number density is $n_{\rm e}=n_{\rm th}+n_{\rm nth}\equiv \rho/(m_{\rm p}+m_{\rm e})$ ($=n_{\rm p}$, the total proton number density, due to charge neutrality), we can simplify the efficiency as,
\begin{equation}
    \epsilon_{\rm PIC} \equiv \frac{\int^{\gamma_{\rm max}}_{\gamma_{\rm min}}(\gamma - 1)\frac{dN_{\rm nth}}{d\gamma}d\gamma}{\int^{\infty}_{\gamma_{\rm pk}}(\gamma - 1)\frac{dN_{\rm th}}{d\gamma}d\gamma+\int^{\gamma_{\rm max}}_{\gamma_{\rm min}}(\gamma - 1)\frac{dN_{\rm nth}}{d\gamma}d\gamma},
    \label{eqn:efficiency}
\end{equation}{}
\noindent where $\gamma_{\rm pk}$ is the peak Lorentz factor of the Maxwellian distribution. To simplify the thermal case, we have chosen the minimum limit of the integration over $dN_{th}/d\gamma$ to be $\gamma=1$ rather than $\gamma_{\rm pk}$, which gives a simpler analytical form for the total thermal energy density shown in Eqn.~\eqref{eqn:thermal_energy_density}. This assumption for the integral limits for the thermal electron energy density results in a larger population of electrons being accelerated to a power-law in high temperature regions as compared to what we expect from Eqn.~\eqref{eqn:efficiency}. However, note that Eqn.~\eqref{eqn:efficiency_ball} is not strictly the same as Eqn.~\eqref{eqn:efficiency} since the numerator in Eqn.~\eqref{eqn:efficiency_ball} is the non-thermal contribution to the electron energy density after the Maxwellian component is removed whereas the numerator in Eqn.~\eqref{eqn:efficiency} includes all electrons with energies $\gamma>\gamma_{\rm pk}$. Hence, the efficiency should be larger than that predicted by the \citet[][]{Ball_2018_PIC} prescription. However, both the inaccuracies mentioned above are small and counteract each other, so we expect errors in the output spectrum to be negligible. Therefore, using Eqn.~\eqref{eqn:nth_energy_density}, the final form of the total non-thermal energy density is as follows: 

\begin{equation}
    U_{\rm nth} = \frac{\epsilon_{\rm PIC}}{1-\epsilon_{\rm PIC}}\left(U_{\rm th}-n_{\rm th}m_{\rm e}c^2\right)+n_{\rm nth}m_{\rm e}c^2
    \label{eqn:efficiency_final}
\end{equation}

\noindent The \citet{Ball_2018_PIC} acceleration efficiency prescription is given as follows:
\begin{equation}
    \epsilon_{\rm PIC}=A_{\epsilon}+B_{\epsilon} \tanh{C_{\epsilon}\beta_{\rm P, amb}},
\label{eqn:efficiency_fit}
\end{equation}

\noindent where $A_{\epsilon}=1-(4.2\sigma_{\rm M, amb}^{0.55}+1)^{-1}$, $B_{\epsilon}=0.64\sigma_{\rm M, amb}^{0.07}$ and $C_{\epsilon}=-68\sigma_{\rm M, amb}^{0.13}$. This fit for the efficiency goes to zero for $\sigma_{\rm M, amb}\ll 1$ (non-relativistic reconnection), and 1 for $\sigma_{\rm M, amb}\gg 1$ (ultra-relativistic reconnection). Figure~\ref{fig:edf} shows a comparison of the electron distribution functions between the two non-thermal acceleration models and clearly shows the broken power-law for a region in the accretion flow close to the BH. From the figure, we see that model $\epsilon_{\rm PIC}$ predicts a steep power-law index and low acceleration efficiency in the inner accretion flow.  

\begin{figure*}  
\begin{centering}
\vspace{-10pt}
    \includegraphics[width=\textwidth,trim=0cm 0cm 0cm 0cm,clip]{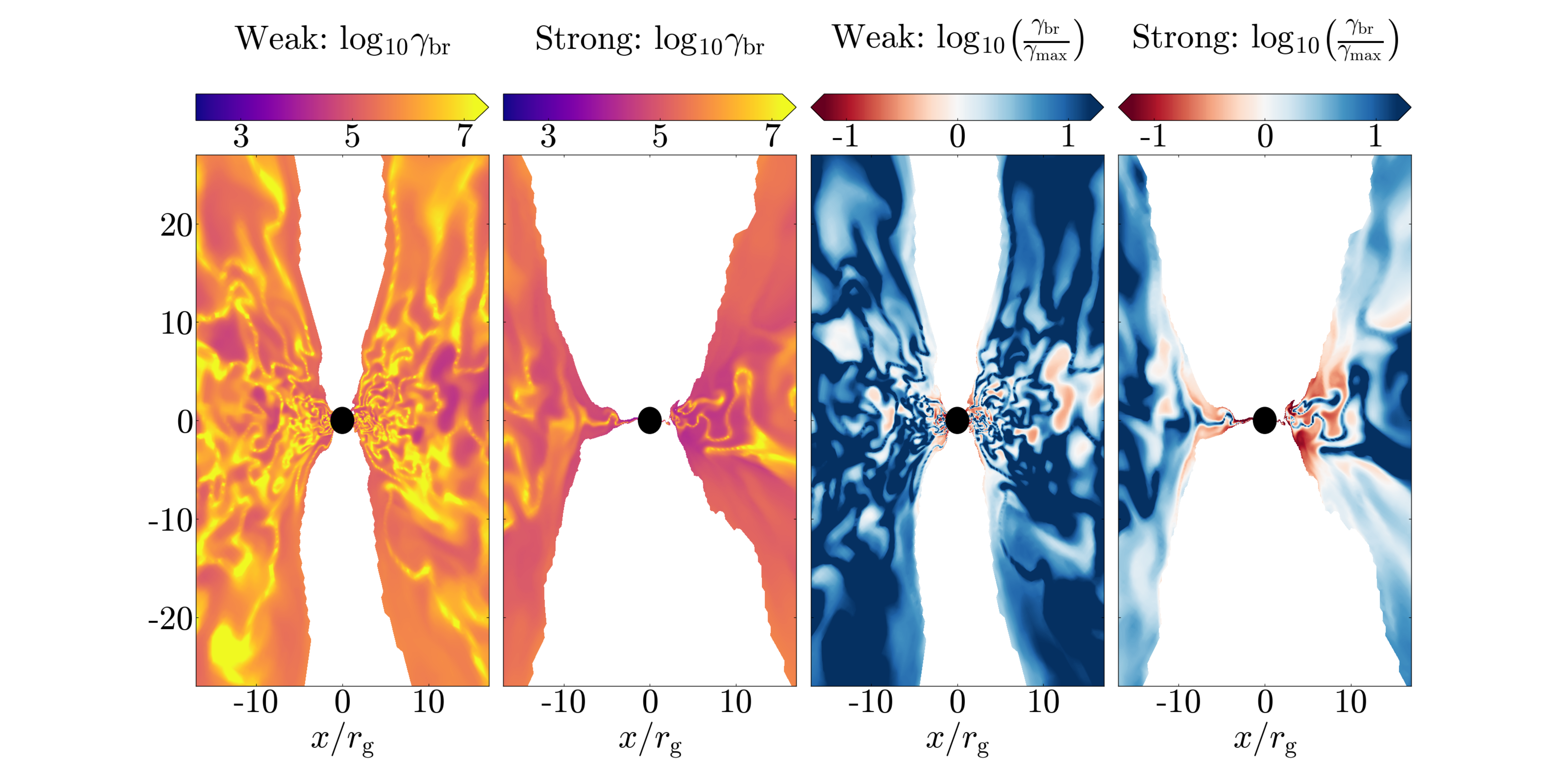}
    \caption{2D cross-sections of the break Lorentz factor $\gamma_{\rm br}$ as determined from Eqn.~\eqref{eqn:gamma_br} and the ratio of $\gamma_{\rm br}$ and the power-law distribution Lorentz factor cutoff $\gamma_{\rm max}$. Most of the synchrotron cooling occurs within $5-10~r_{\rm g}$ of the BH, where $\gamma_{\rm br} < \gamma_{\rm max}$. We take the same snapshots for this figure as the ones shown in Fig.~\ref{fig:eff_pl}. As in Fig.~\ref{fig:eff_pl}, here we exclude the jet funnel from our calculations.}
    \label{fig:nnth_gbr}
\end{centering} 
\end{figure*}  

\section{Results}
\label{sec:results}

\subsection{GRMHD evolution}
\label{sec:grmhd_results}

\begin{figure}  
\begin{centering}
\vspace{-10pt}
    \includegraphics[width=\columnwidth,trim=0cm 0cm 0cm 0cm,clip]{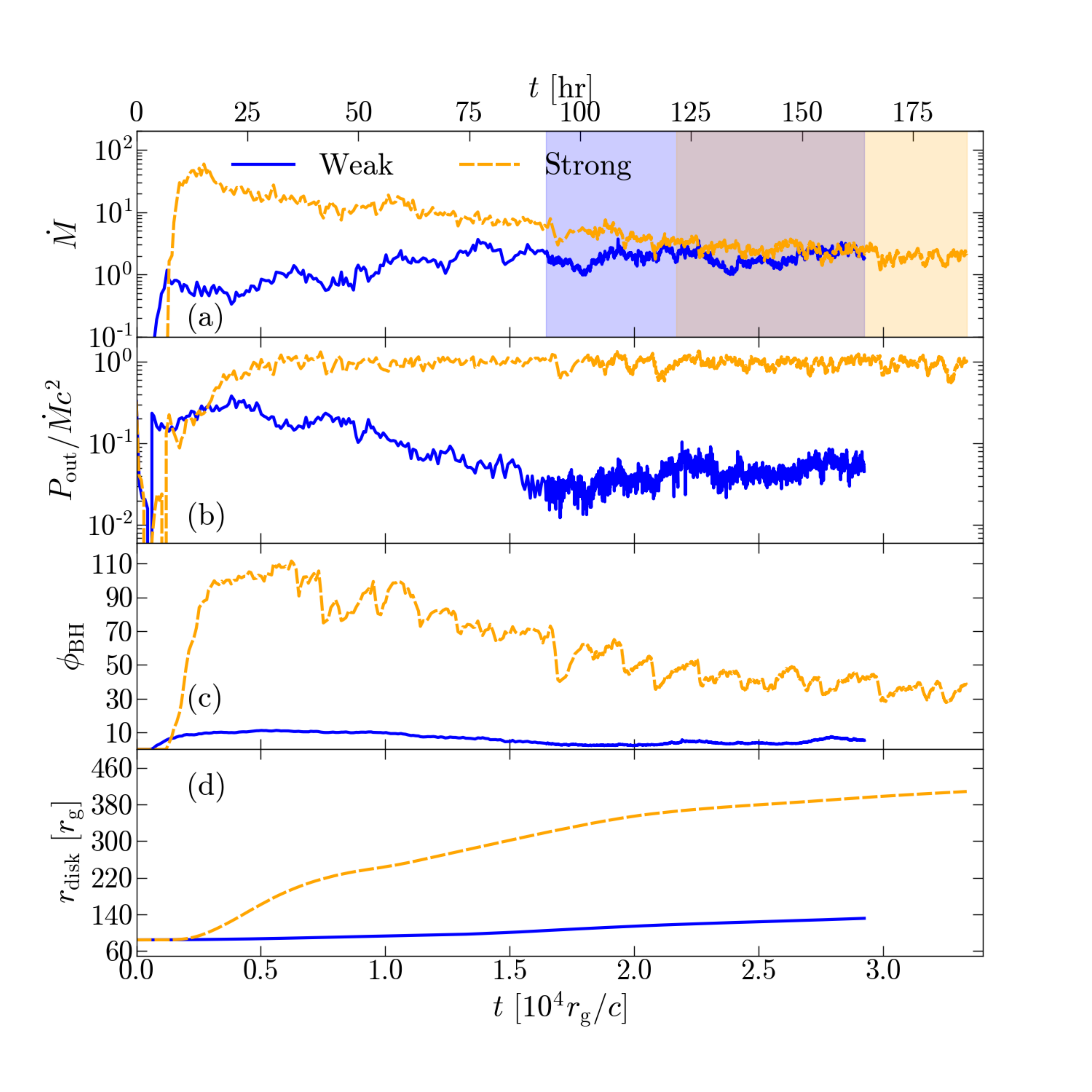}
    \caption{Time evolution of several simulation physical parameters, comparing the \weak{} and \strong{} models. Panels show: (a) mass accretion rate $\dot{M}$ in $M_{\odot}$~yr$^{-1}$, (b) dimensionless outflow power efficiency $P_{\rm out}/\dot{M}c^2$, (c) dimensionless magnetic flux $\phi_{\rm BH}$ in Gaussian units, and (d) barycentric radius $r_{\rm disc}$ in units of $r_{\rm g}$. We measure all quantities at the event horizon. Section~\ref{sec:grmhd_results} lists the definition of each quantity. We ray-trace each simulation over the corresponding shaded time segment (blue for \weak{} and orange for \strong{}).}
    \label{fig:time}
\end{centering} 
\end{figure} 

We evolve model \weak{} to $2.93\times 10^4 r_{\rm g}/c$ or $164$ hours for \sgra{}, and model \strong{} to $3.34\times10^4 r_{\rm g}/c$ or $187$ hours for \sgra{}. Figure~\ref{fig:grmhd} shows 2D cross-sections of the electron number density and temperature as well as the plasma-$\beta$ of our GRMHD models, scaled to the parameters of our radiative models for \sgra{} (see Table~\ref{tab:models}). The overall jet and disc properties are significantly different between the two models, with the strong-field disc model \strong{} displaying wider outflows and lower disc electron number densities and temperatures. The low plasma-$\beta$ and high temperatures in the \strong{} outflow region suggest that the bulk of the synchrotron emission in the sub-millimetre waveband for \sgra{} would originate in the jet sheath, whereas for the weak-field disc model \weak{}, the disc would produce a dominant fraction of the radiation. This result crucially depends on the electron temperature prescription shown in Sec.~\ref{sec:GRRT_setup} \citep[also see Fig.~4 in][]{EHTPaperV}. 

Given that the non-thermal acceleration efficiency and power-law index is sensitive to the local magnetisation and plasma-$\beta$, we expect that regions of low plasma-$\beta$, i.e., the jet sheath, would dominate the majority of X-ray emission. Figure~\ref{fig:eff_pl} shows that the acceleration efficiency is indeed only significantly high in the jet sheath for the \weak{} model while for the \strong{} model, the extended jet sheath and the low plasma-$\beta$ near the BH both exhibit non-zero non-thermal acceleration. The individual values of the efficiency in the two non-thermal models ($\epsilon_{\rm C}$ and $\epsilon_{\rm PIC}$) differ greatly due to the model definitions, i.e., one is defined as a fraction of the magnetic energy density and the other as a fraction of the electron energy density. Further, for model $\epsilon_{\rm PIC}$, we only get power-law indices less than four at the jet boundary. Thus, we see that non-thermal electrons only appear at the jet-edge in $\epsilon_{\rm PIC}$ whereas there is a wider particle acceleration region in $\epsilon_{\rm C}$ models. Further, Fig.~\ref{fig:nnth_gbr} shows that the synchrotron cooling only occurs in parts of the disc and the jet sheath, where the break Lorentz factor is smaller than the power-law cutoff Lorentz factor $\gamma_{\rm max}$. The spectral steepening would thus affect the X-ray emission which originates in the current sheets close to the BH.

Figure~\ref{fig:time} shows the time evolution of our simulations, illustrating the horizon accretion rate $\dot{M}$, the horizon energy accretion rate $\dot{E}$, the horizon outflow efficiency (i.e., the ratio of the outflow power $P_{\rm out}$ and the accretion power $\dot{M}c^2$), the horizon dimensionless magnetic flux $\phi_{\rm BH}=\Phi_{\rm BH}/(\langle\dot{M}\rangle r_{\rm g} c^2)^{1/2}$ (in Gaussian units) and the disc barycentric radius $r_{\rm disc}$, defined as:
\begin{equation}
    \dot{M}=-\iint \rho u^r \, \! \sqrt{-g}\, d\theta \, d\varphi \,,
\end{equation}
\begin{equation}
    \dot{E}=\iint T^r_t \, \! \sqrt{-g}\, d\theta \, d\varphi \,,
\end{equation}
\begin{equation}
    P_{\rm out}=\dot{M}c^2-\dot{E}\,,
\end{equation}
\begin{equation}
    \Phi_{\rm BH}=\frac{\sqrt{4\pi}}{2}\iint |B^{r}| \, \sqrt{-g}\, d\theta \, d\varphi\,,
\end{equation}{}
\begin{equation}
    r_{\rm disc}=\dfrac{\iiint r \, \rho \,  \sqrt{-g}\, dr\, d\theta \, d\varphi}{\iiint \rho \, \sqrt{-g}\, dr\, d\theta \, d\varphi}\,.
    \label{eqn:barycentric_radius}
\end{equation}
where $T^r_t$, $u^r$, $B^r$ and $g\equiv|g_{\mu\nu}|$ are the total radial energy flux, radial velocity, radial magnetic field and the metric determinant respectively \citep[standard definitions from e.g., ][]{Porth:19,Chatterjee_2020}. The weak-field model, \weak{}, produces a jet with an average efficiency of $P_{\rm out}/\dot{M}c^2 \lesssim 6\%$ while the strong-field model (\strong{}) jet attains an efficiency of almost $100\%$. Jet efficiencies $\gtrsim100\%$ are known to occur when the magnetic pressure around the BH becomes sufficiently dominant to obstruct gas from accreting \citep[e.g.,][]{narayan03, tch11}, leading to a magnetically arrested disc (MAD) state. Indeed, as shown in \citet{Dexter:2020:NIR_MAD} and \citet{Porth:2020:NIR_MAD}, MAD conditions lead to magnetic flux eruptions where magnetised low density bunches of field lines (i.e., flux tubes) escape from the BH's event horizon and interact with the surrounding accretion material via shearing instabilities, heating up electrons and causing NIR flaring events \citep{Dexter:2020:NIR_MAD}. 

Neither simulation develops the MAD state as they do not exhibit magnetic flux eruptions. However, \strong{} displays a significantly higher jet power as well as horizon magnetic flux, which considerably changes disc-jet dynamics close to the BH, especially with respect to disc turbulence. Stronger outflows lead to a higher rate of momentum transport outwards and hence causes the disc to viscously spread out as illustrated by the increasing value of $r_{\rm disc}$ over time in the case of \strong{}. One important question remains for MADs or even strong-field discs in the context of \sgra{}: these types of discs always produce powerful jets that seem to be absent in \sgra{}. From the ongoing EHT observations of \sgra{}, we will hopefully be able to place much better constraints on the size of the source as well as the structure of the horizon-scale flow, which will prove crucial to settling the question of jets from \sgra{}. Our goal is to understand the effect of disc or jet turbulence on variability in a quasi-stable disc whereas for MADs, the presence of magnetic eruptions distorts the inner structure of the accretion flow.

\begin{figure*} 
\begin{centering}
    \includegraphics[width=\textwidth,trim=0cm 0cm 0cm 0cm,clip]{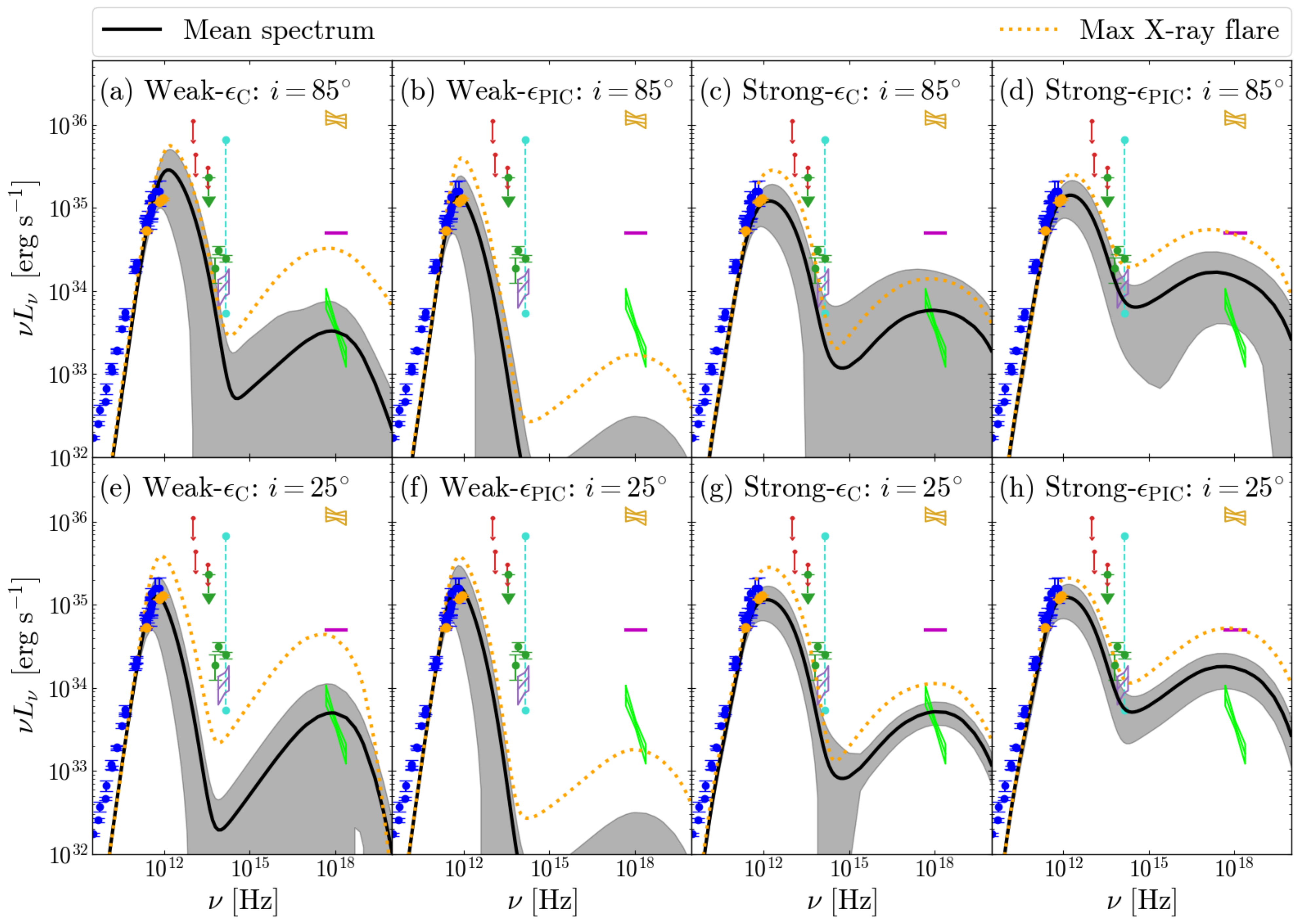}
    \caption{Multiwavelength synchrotron-only spectra for each radiative model at two inclination angles: \weakC{} - weak-field simulation $+$ constant power-law injection non-thermal model, \weakPIC{} - weak-field simulation $+$ varying power-law injection non-thermal model, \strongC{} - strong-field simulation $+$ constant power-law injection non-thermal model and \strongPIC{} - strong-field simulation $+$ varying power-law injection non-thermal model, for two inclination angles: (top row) $85^{\circ}$ and (bottom row) $25^{\circ}$. We show the mean spectrum over the considered time segment (black solid line) along with $1\sigma$ standard deviations from mean (in grey), and the spectrum for the simulation snapshot with the brightest X-ray flare (orange dotted line). We include radio and sub-mm data points quoted in \citet[][blue circles]{Connors_2017}, compiling historical data from \citet{Zylka_1995, Serabyn_1997, Falcke_1998, Zhao_2001, Nord_2004, Roy_2004, An_2005, Lu_2011, Brinkerink_2015, Bower_2015}. We also include sub-mm observations at 233, 678 and 870~GHz from \citet[][orange circles]{Bower_2019}, with mid-infrared datapoints from \citet[][green circles and an arrow]{Schodel_2011} and infrared upper bounds from \citet[][red arrows]{Melia_2001}. Furthermore, we combined the \citet{Gravity:2020:flux} median flux $1.1\pm 0.3$~mJy at $2.2~\mu$m with the spectral index $\alpha_{\nu}=-0.6\pm 0.2$ measured over $1.6-3.8~\mu$m from \citet{Hornstein_2007} to create the purple bowtie. We show the maximum and minimum dereddened fluxes ($59.6$ and $0.48$~mJy, respectively) at 2.2$\mu$m from \citet[][cyan circles]{Do_2019} to guide our NIR flare spectra. For the X-ray quiescent spectrum (green bowtie), we take the 2-8~keV luminosity of $L_{\rm X}=(3.6\pm 0.4)\times 10^{33}$~erg s$^{-1}$ and photon index $\Gamma=3.0\pm 0.2$ from \citet{Nowak_2012}, while for the flaring state, we show the average flare spectrum with $L_{\rm X}\simeq 5\times 10^{34}$~erg s$^{-1}$ with $\Gamma=2$ from \citet[][magenta line]{Neilsen_2015}. We also show the brightest X-ray flare detected to date: a double-peaked flare with $L_{\rm X}=(12.26^{+0.28}_{-0.27}, 10.97^{+0.28}_{-0.27})\times 10^{35}$~erg s$^{-1}$ and $\Gamma=2.06\pm 0.1$ from \citet[][golden bowtie]{Haggard_2019}. There is a significant difference in synchrotron emission from the considered hybrid thermal$+$non-thermal electron energy distributions, particularly in variability when comparing the weak and strong-field models. None of the models achieve the high NIR and X-ray luminosities seen in \citet{Do_2019} and \citet{Haggard_2019}, respectively.}
    \label{fig:spectra}
    \end{centering}
\end{figure*}

\subsection{Multiwavelength spectrum of \sgra{}: observations and GRRT modelling}
\label{sec:sed}

We generate synchrotron spectra for each of our hybrid thermal$+$non-thermal models assuming the BH mass and distance of \sgra{} (see Table~\ref{tab:models}: GRRT parameters). From Fig.~\ref{fig:time}(a), we see that the accretion rate for each model at first increases, marking the start of accretion from the torus, and then slowly decreases and finally settles at a time $t\simeq 1.5\times 10^4 r_{\rm g}/c$ for model \weak{} and at $t\simeq 2\times 10^4 r_{\rm g}/c$ for model \strong{}. We choose a time segment spanning in excess of 60 hours for each simulation. The time duration is quite long compared to the dynamical timescales for typical flares in \sgra{}, which are on the order of minutes to an hour. The fiducial model source inclination angle with respect to the observer is taken to be $85^{\circ}$ in accordance with previous \sgra{} models \citep[e.g.,][]{Markoff_2007,Moscibrodzka_2009,Shcherbakov_2012,Drappeau_2013,Connors_2017}. \sgra{}'s inclination angle is still an open question with several other works employing smaller inclinations \citep[e.g.,][]{Dexter_2010,moscibrodzka_2013,Chael:2017,Davelaar2018,Gravity:20:orbit}. Hence, we also perform the same GRRT calculations at an inclination angle of $25^{\circ}$. We do not change any other parameter, such as the electron temperature distribution or the accretion rate, in order to directly compare to the fiducial $85^{\circ}$ inclination models. The source azimuthal angle with respect to the observer is taken to be $0^{\circ}$ as the BH disc-jet system is roughly axisymmetric \citep[this quantity only becomes important for misaligned BH discs, see][]{Chatterjee_2020}. The field-of-view (FOV) for the GRRT imaging is $50~r_{\rm g}\times 50~r_{\rm g}$ with an image resolution of $1024\times 1024$~pixels. 

Figure~\ref{fig:spectra} shows the multiwavelength spectrum for each radiative model: the mean spectrum along with $1\sigma$ deviations, and the maximum X-ray flare spectrum. The mass accretion rate for each model is chosen to match the observed sub-mm fluxes: $3.69\times 10^{-8}\,M_{\odot}$~yr$^{-1}$ for \weak{} and $2.12\times 10^{-8}\,M_{\odot}$~yr$^{-1}$ for \strong{}. These accretion rates lie well within the range of $2\times 10^{-9}~M_{\odot}$~yr$^{-1} < \dot{M}<2\times 10^{-7}~ M_{\odot}$~yr$^{-1}$, inferred from Faraday rotation measurements of \sgra{} \citep{Bower:03,Marrone:07}. The choice of $R_{\rm high}$ and $R_{\rm low}$ is made such that the mean spectrum roughly fits the 230~GHz flux while not overproducing the NIR emission. We note that in the case of \weakPIC{} (Fig.~\ref{fig:spectra}b), lower values of $R_{\rm low}$ can be used to preferentially heat the electrons in low plasma-$\beta_{\rm P}$ regions, such as the jet sheath. However, for this work, to keep the models comparable, we choose the same value of $R_{\rm low}$ for each model, varying only $R_{\rm high}$ and therefore the electron temperature in the disc. The change in spectral index at frequencies above $10^{17}$~Hz occurs due to synchrotron cooling, with the turnover frequency set by equating the local advective and synchrotron cooling timescales. We reiterate at this point that we do not account for inverse Compton scattering, and the $10^{17}$~Hz bump is due to non-thermal synchrotron only.

During periods of flaring, the non-thermal contribution to the NIR flux increases and affects the NIR slope in the spectra of three models (Fig.~\ref{fig:spectra}a,b,d), while in the case of \strongC{}, the thermal synchrotron emission dominates the NIR emission. None of the models are able to simultaneously reproduce the NIR \textit{GRAVITY} flux and the \citet{Hornstein_2007} slope, while the brightest X-ray flare spectrum of \weakC{} displays a similar slope, but fails to produce the required NIR flux. In the X-rays, \weakC{} is consistent with the quiescent X-ray flux while \weakPIC{} displays a low X-ray quiescent flux. The mean spectrum for the strong-field models overproduce the quiescent X-ray limits. Since the inner disc in \strong{} is highly magnetised, we see higher X-ray emission overall with model \strongPIC{} achieving the brightest flare with an X-ray luminosity exceeding the average flare luminosity. Among the models that produce X-ray luminosities above quiescence, \weakC{} produces the largest relative change in X-ray luminosity between the mean and brightest flare spectrum with a difference of over an order of magnitude. It is further encouraging to note that in the case of the PIC-motivated radiation models, the change in power-law index $p$ of the non-thermal distribution occurs at $\nu\sim 10^{17}$~Hz due to synchrotron cooling and hence, the X-ray power-law index during the brightest flares is $p+1 \sim 3$, consistent with the photon indices measured by \citet{Nowak_2012} and \citet{Haggard_2019}. This result verifies the claim that simultaneous NIR and X-ray flares are perhaps related via a broken power-law distribution \citep[e.g.,][]{Dodds-Eden_2010, Dibi_2014, Ponti_2017}, and strongly points towards a non-thermal origin for \sgra{} flaring. We note that the synchrotron spectra from our models do not display a clean cooling break as seen in semi-analytical models such as \citet[][]{Dodds-Eden_2010,Ponti_2017}. This difference occurs because our spectra are a summation of power-law emission from different regions in the flow while these papers use a single-zone emission model with a single broken-power-law component.

The $25^{\circ}$ inclination models show similar mean spectra to their $85^{\circ}$ inclination counterparts. If we compare the \weakC{} models (panels a and e), the NIR emission drops for lower inclination. This reduction occurs because the NIR emission originates in the toroidal gas flow in the accretion flow, which is Doppler-boosted towards us at high inclinations. This effect is less evident in the \strong{} models since the primary source of NIR photons is the broad jet sheath. The X-ray spectra do not change significantly when changing the inclination angle since this is optically thin synchrotron emission. We will discuss variability in the low inclination case in the next section.

\subsection{Disc-averaged profiles}
\label{sec:radial_profiles}

\begin{figure}  
\begin{centering} 
    \includegraphics[width=\columnwidth,trim=0cm 0cm 0cm 0cm,clip]{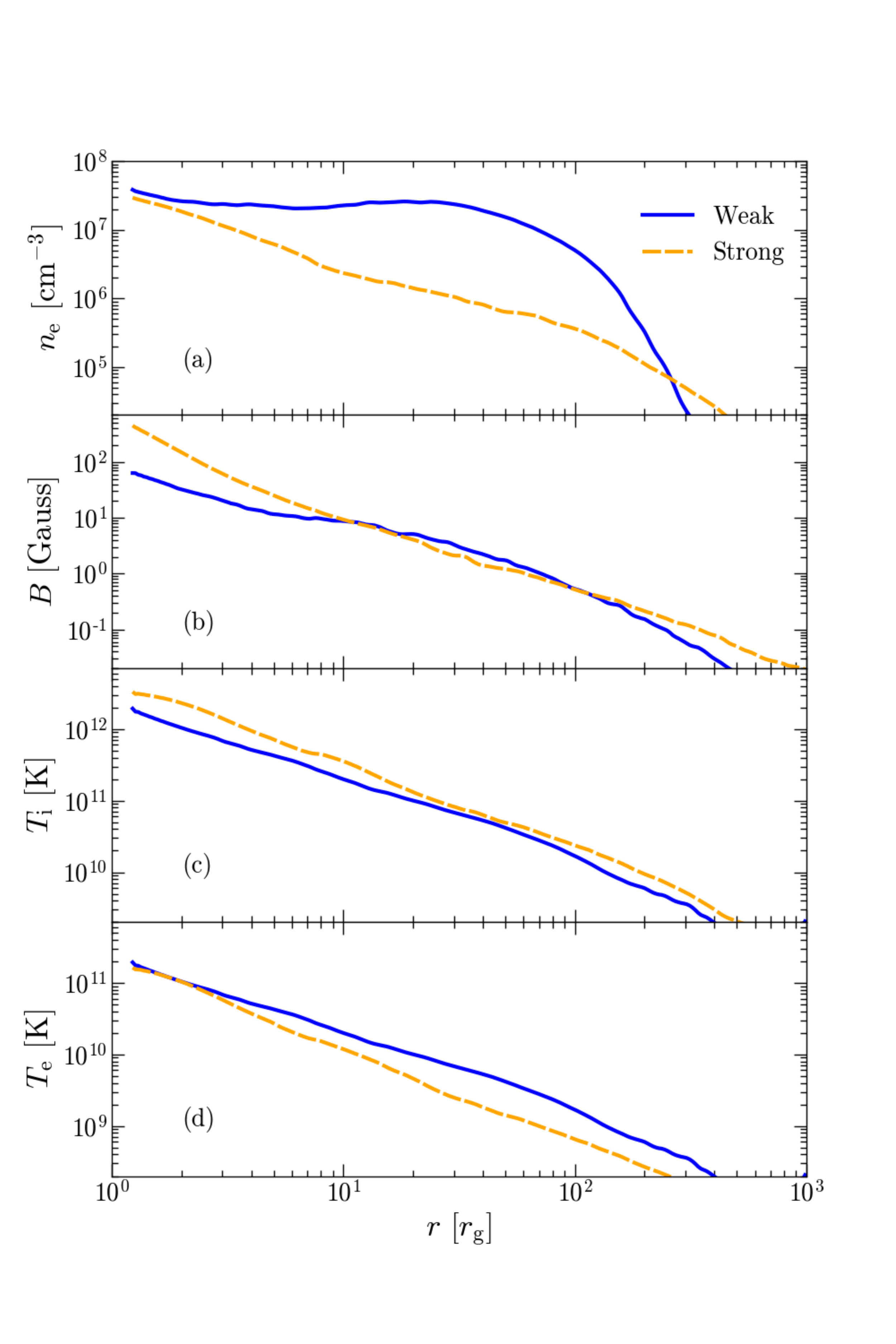}
    \caption{Radial dependencies of the disc-averaged quantities at $t\sim 28,000~r_{\rm g}/c$: (a) electron number density $n_{\rm e}$ in cm$^{-1}$, (b) the magnetic field strength $B$ in Gauss, (c) the ion (or proton) temperature $T_{\rm i}$ and (d) the electron temperature $T_{\rm e}$ in Kelvin, when scaled to the mass and accretion rate of \sgra{}. We calculate $T_{\rm e}$ using eqns.~\eqref{eq:R_parameter} and \eqref{eq:elec_temp} as per quoted values in Table~\ref{tab:models}. \weak{} contains a larger electron concentration in the disc and exhibits a near constant radial profile, while \strong{} displays a steeper power-law profile due to stronger disc turbulence. The magnetic field strengths are similar between the two models, apart from the inner $5~r_{\rm g}$ where \strong{} displays field strengths larger by factors of 5-8. Stronger turbulence leads to slightly higher ion temperatures in the inner disc of \strong{}, whereas the electron temperatures in the two models behave very similarly, exhibiting an approximate $r^{-1}$ profile.}
    \label{fig:radial}
\end{centering} 
\end{figure}  

Figure~\ref{fig:radial} shows the disc-averaged values of the electron number density $n_{\rm e}$ in cm$^{-3}$, magnetic field strength $B$ in Gauss, and the ion and electron temperatures, $T_{\rm i}$ and $T_{\rm e}$ respectively, in Kelvin for both \weak{} and \strong{} simulations, when scaled to the BH mass of \sgra{} and the aforementioned accretion rates. The disc-averaging for a parameter $q\in (\rho, B, T_{\rm i}, T_{\rm e})$ is calculated in the form
\begin{equation}
    <q>=\dfrac{\iint q \, \rho \,  \sqrt{-g}\, d\theta \, d\varphi}{\iint \rho \, \sqrt{-g}\, d\theta \, d\varphi}\,,
\end{equation}{}

\noindent similar to the evaluation of the barycentric radius $r_{\rm disc}$ in the previous section. As mentioned in the previous section, from Fig.~\ref{fig:time}(d), we see that the strong-field \strong{} disc grows larger over time and hence, becomes more diffused, leading to lower $n_{\rm e}$ values in the disc as compared to the weak-field model \weak{} (Fig.~\ref{fig:radial}a). Even though the \strong{} ion temperature is higher, the electron temperatures from the two simulations are roughly similar in the disc since the $R_{\rm high}$ parameter for \strong{} is 4 times as large as for \weak{}. In the next section, we study the variable properties of the observed lightcurves that are the outcome of disc turbulence and magnetic reconnection in the current sheets.

\section{Variability in lightcurves}
\label{sec:variability}

\begin{figure*} 
\begin{centering}
    \includegraphics[width=\textwidth,trim=0cm 0cm 0cm 0cm,clip]{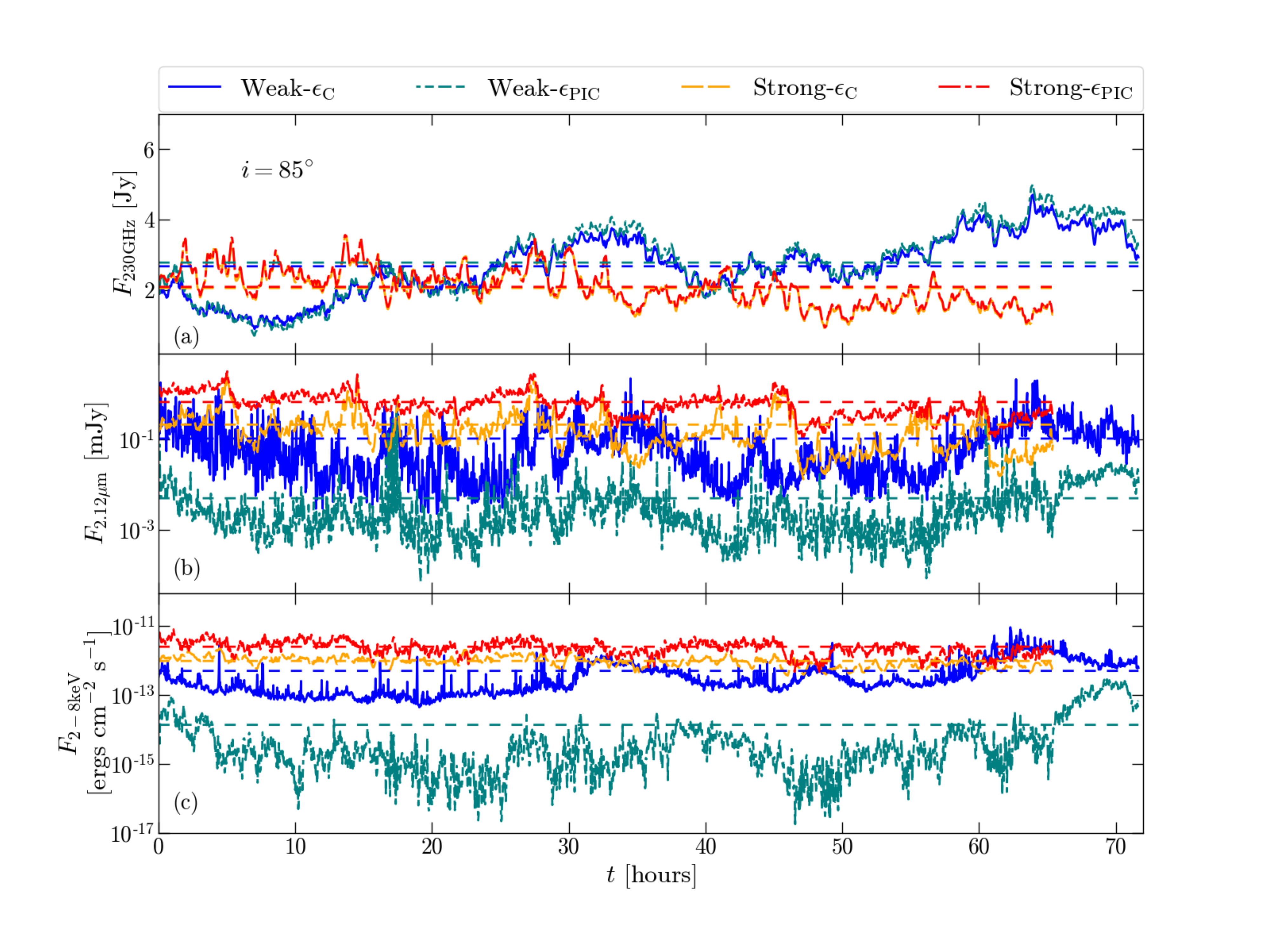}
    \caption{GRRT lightcurves for the $85^{\circ}$ inclination radiative models at 3 wavebands in the standard units used in the literature: (a) 230~GHz in Jansky [Jy], (b) near-infrared (NIR at $2.12~\mu$m) in milli-Jansky [mJy] and (c) X-rays (integrated over 2-8~keV) in erg cm$^{-2}$ s$^{-1}$. The 230~GHz lightcurves largely follow the corresponding model accretion rate over time and lightcurves from a particular simulation model (\weak{} or \strong{}) behave in a similar fashion as the same thermal electron population dominates the 230 GHz emission in each respective GRMHD model. The NIR lightcurves for the \strong{} radiative models are also roughly correlated since thermal synchrotron emission dominates the NIR emission. For \weakC{}, both thermal and non-thermal electrons contribute to the NIR emission, whereas there is hardly any significant non-thermal contribution in \weakPIC{}. The lack of non-thermal electrons in \weakPIC{} results in extremely low X-ray fluxes while \weakC{} exhibits intermittent flares. The X-ray lightcurve for \strongPIC{} looks more variable with relatively larger changes in flux as compared to \strongC{}.}
    \label{fig:lightcurves}
    \end{centering}
\end{figure*}

In this section, we perform timing analysis of the radiative models studied in the previous section. We consider three specific wave bands and compare our results with three corresponding observational papers: 230~GHz - \citet{Dexter_2014}, 2.12~$\mu$m - \citet{Do_2019} and 2-8~keV - \citet{Neilsen_2015}. Figure~\ref{fig:lightcurves} shows the lightcurves in each of the three bands at inclination $85^{\circ}$, for each model (Table~\ref{tab:models}). For comparison with the observational data, we construct the fractional root-mean-square (rms) normalised power spectral density (PSD) curves for each set of lightcurves (Fig.~\ref{fig:rms}) and cumulative distribution functions (CDFs) for the 2.2~$\mu$m and X-ray lightcurves (Figs.~\ref{fig:NIR_CDFs}, \ref{fig:Xray-cdf-85} and \ref{fig:Xray-cdf-25}). To reduce noise at high frequencies in the PSDs, we re-bin logarithmically and average over frequency bins. In the following subsections, we look at each waveband in turn and discuss the lightcurves, PSDs and CDFs for each $85^{\circ}$ inclination model and then compare to the corresponding $25^{\circ}$ inclination case. 

\begin{figure*} 
\begin{centering}
    \includegraphics[width=\textwidth,trim=0cm 0cm 0cm 0cm,clip]{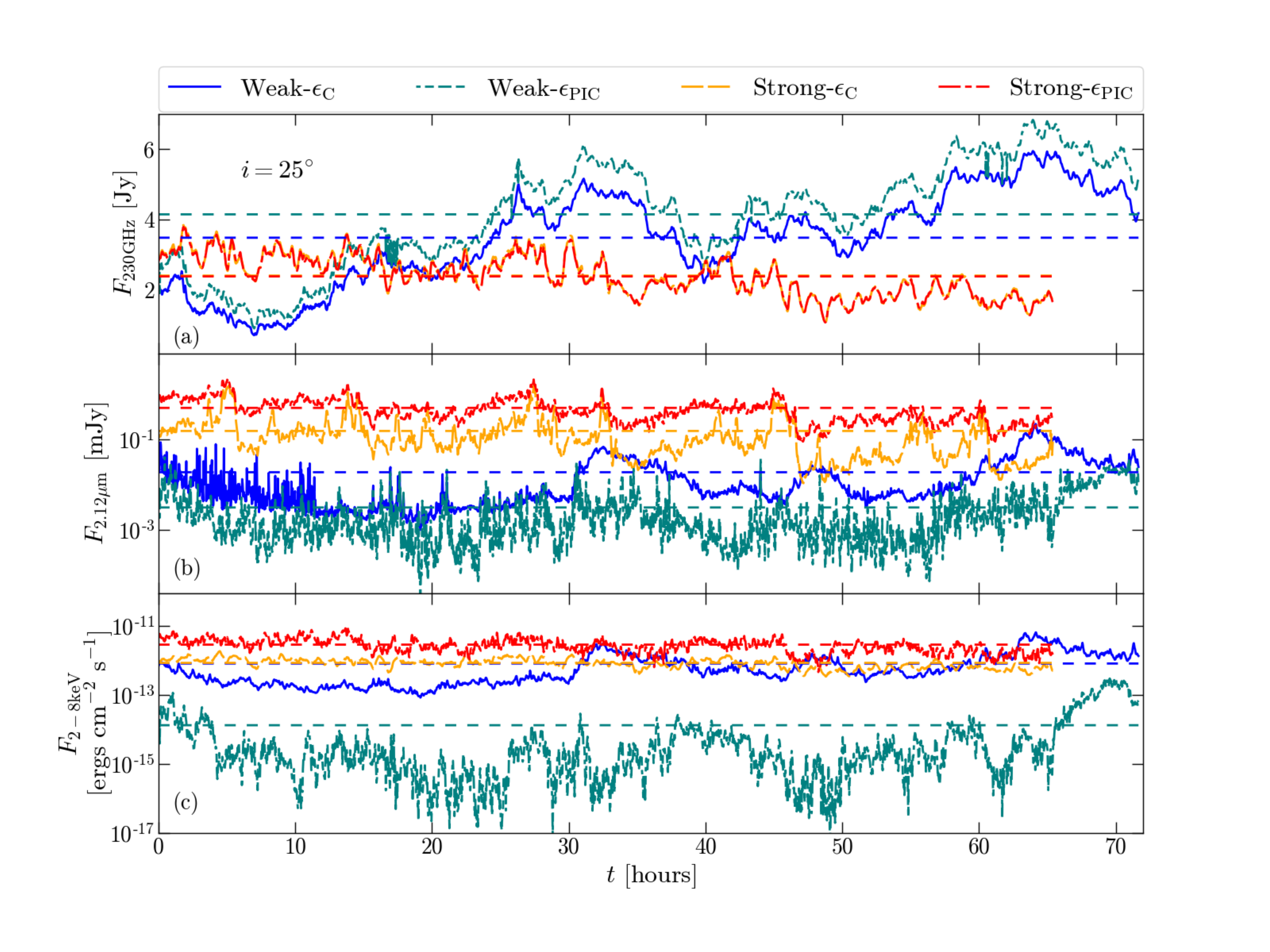}
    \caption{Same as Fig.~\ref{fig:lightcurves}, but for an inclination angle of $25^{\circ}$. Overall, the lightcurves look similar to their $85^{\circ}$ counterparts. The 230~GHz mean flux increased by a factor of less than 2 when imaging at $25^{\circ}$. The \weakC{} NIR lightcurve also seems to appear less noisy when compared to its $85^{\circ}$ counterpart.}
    \label{fig:lightcurves_25}
    \end{centering}
\end{figure*}

\subsection{Sub-millimetre: 230~GHz}
\label{sec:submm}

Figure~\ref{fig:lightcurves}(a) shows that the weak-field model 230~GHz lightcurves- \weakC{} and \weakPIC{} are similar to each other on average and show an increase of a factor $\lesssim 2$ in the flux at late times, largely following the variations in the accretion rate. For the strong-field disc cases, \strongC{} and \strongPIC{}, the lightcurves behave quite similarly: the 230~GHz lightcurves lie almost on top of each other. The similarity in the lightcurves suggests that the bulk of the 230~GHz emission is being produced in the same region for each GRMHD model, as is expected since the emission is thermal in nature and there is only one electron temperature distribution for each simulation model. The accretion rate is scaled such that the 230~GHz flux is the same for all models. However, as seen from the average spectrum in Fig.~\ref{fig:spectra}(a) vs (b), there is a noticeable addition to the terahertz flux in the case of model \weakC{} due to the presence of a higher percentage of thermal electrons in the jet sheath in comparison to \weakPIC{}. This difference in the terahertz flux is due to the acceleration prescription used in \weakC{} where magnetic energy is transferred to the non-thermal energy density. Hence, the regions closest to the BH (where the magnetic field is the strongest) is favoured as the region of electron acceleration as opposed to the jet sheath. However, for the strong-field radiative models, the radio-to-infrared spectra are almost similar. The $25^{\circ}$ inclination 230~GHz lightcurves in Fig.~\ref{fig:lightcurves_25}(a) display a higher flux as compared to the $85^{\circ}$ inclination models as the image becomes more extended at low inclinations. 

\begin{figure}  
\begin{centering}
    \includegraphics[width=\columnwidth,trim=0cm 0cm 0cm 0cm,clip]{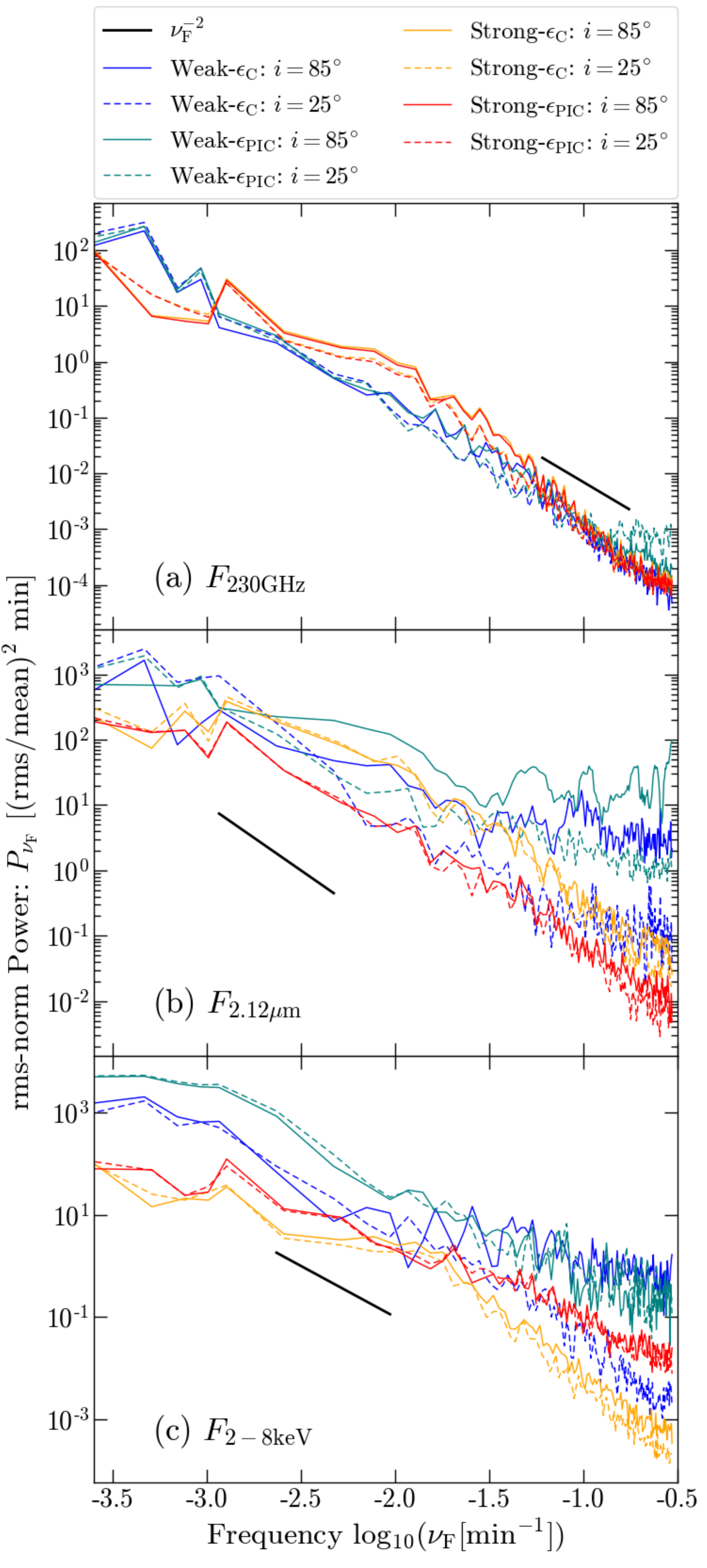}
    \caption{Fractional root-mean-square (rms) normalised power spectral density (PSD; $P_{\nu_{\rm F}}$) plots for each radiative model at (a) 230~GHz, (b) $2.12~\mu$m and (c) 2-8~keV integrated X-ray flux. The x-axis frequency $\nu_{\rm F}$ corresponds to the inverse of timescales. Solid and dashed lines indicate $85^{\circ}$- and $25^{\circ}$ inclination models respectively. We also show a black line in each plot representing $\nu_{\rm F}^{-2}$ dependence, a characteristic of red noise turbulence. At high frequencies, while the 230~GHz PSDs behave similar to red noise, the NIR and X-ray PSDs exhibit shallower profiles at high frequencies, closer to $\nu_{\rm F}^{-1}$ and even frequency-independent behaviour. The absence of pronounced white noise at low frequencies indicates that we are unable to capture long-timescale variability due to the short length of our lightcurves. $25^{\circ}$ inclination models largely behave similar to their $85^{\circ}$ counterparts except for the \weak{} models in the NIR.}
    \label{fig:rms}
\end{centering} 
\end{figure}  

Figure~\ref{fig:rms}(a) shows the 230~GHz PSD as a function of the Fourier sampling frequency ($\nu_{\rm F}$) for each model. At high frequencies, all models behave like red noise with a power-law dependence on $\nu_{\rm F}$ of $\approx -2$. The 230~GHz rms\% value of all the models (Table~\ref{tab:rms_fits}) matches the accretion rate rms\% as well as the observed 20-30\% variation seen in the sub-mm lightcurves of \sgra{} \citep[e.g.,][]{Zhao:2003,Marrone:2008,Dexter_2014}. For the weak-field models, the red-to-white turnover occurs at frequencies close to $0.001$~min$^{-1}$ pointing to a characteristic variability timescale ($\tau_{\rm rms}$) of the order of tens of hours. For the strong-field model, the turnover appears to occur at slightly higher frequencies ($\sim 0.003$~min$^{-1}$), which is close to measured variability timescale of \sgra{} \citep[$\approx 8$~hours of \sgra{};][]{Dexter_2014}. Currently, the lowest frequency bins in the averaged PSDs contain a single datapoint, and hence the standard error for each bin is equal to the power itself, which would introduce large error-bars for the best-fit value of variability timescale. Capturing the white noise regime properly is essential for accurately fitting for the variability timescale, and hence, requires a lightcurve that is at least one order of magnitude longer than that calculated in this study.

\subsection{Near-infrared: $2.12~\mu$m}
\label{sec:NIR}

\begin{table*}
\begin{center}
\renewcommand{\arraystretch}{1.3}
\begin{tabularx}{1.35\columnwidth}{l | c c c | c }
\hline\hline
\vspace*{0mm}
& & Mean flux \& ${\rm rms\%}$ & & $\dot{M}$ \\
Model & 230~GHz & 2.12~$\mu$m & 2-8~keV & ${\rm rms\%}$ \\
\hline
& [Jy \& \%] & [mJy \& \%] & [$10^{-13}$ergs cm$^{-2}$ s$^{-1}$ \& \%] & \\
\hline\hline
Inclination: $85^{\circ}$\\
\hline

\weakC{} & $2.68$, 31.58\%  & $0.106$, 158.70\% & $5.07$, 143.94\% & 24.5\%    \\
\weakPIC{} & $2.78$, 34.76\% & $0.005$, 304.59\% & $0.138$, 275.63\% & 24.5\%   \\
\strongC{} & $2.06$, 25.33\%  & $0.216$, 123.66\% & $9.84$, 33.05\% & 23.8\%   \\
\strongPIC{} & $2.09$, 24.43\% & $0.683$, 61.33\% & $25.48$, 45.04\% & 23.8\%   \\

\hline\hline
Inclination: $25^{\circ}$\\
\hline

\weakC{} & $3.49$, 38.20\%  & $0.019$, 144.00\% & $8.29$, 117.17\% & 24.5\%    \\
\weakPIC{} & $4.15$, 35.34\% & $0.003$, 150.42\% & $0.134$, 293.47\% & 24.5\%   \\
\strongC{} & $2.41$, 23.50\%  & $0.156$, 126.64\% & $8.66$, 31.06\% & 23.8\%   \\
\strongPIC{} & $2.39$, 23.01\% & $0.506$, 60.96\% & $29.31$, 45.04\% & 23.8\%   \\
\hline\hline
\end{tabularx}
\end{center}
\caption{Particle acceleration results in highly variable NIR and X-ray emission for \sgra{}. We show the mean flux of the 230~GHz, 2.12~$\mu$m and 2-8~keV \sgra{} lightcurves and the corresponding fractional rms amplitude (rms\%) from the power spectra (Fig.~\ref{fig:rms}). We also show the accretion rate rms\% for each model. While the 230~GHz rms\% values are uniform across all models, there is a large range of rms\% in both the NIR and the X-ray lightcurves.}
\label{tab:rms_fits}
\end{table*}

\begin{figure}  
\begin{centering}
    \includegraphics[width=\columnwidth,trim=0cm 0cm 0cm 0cm,clip]{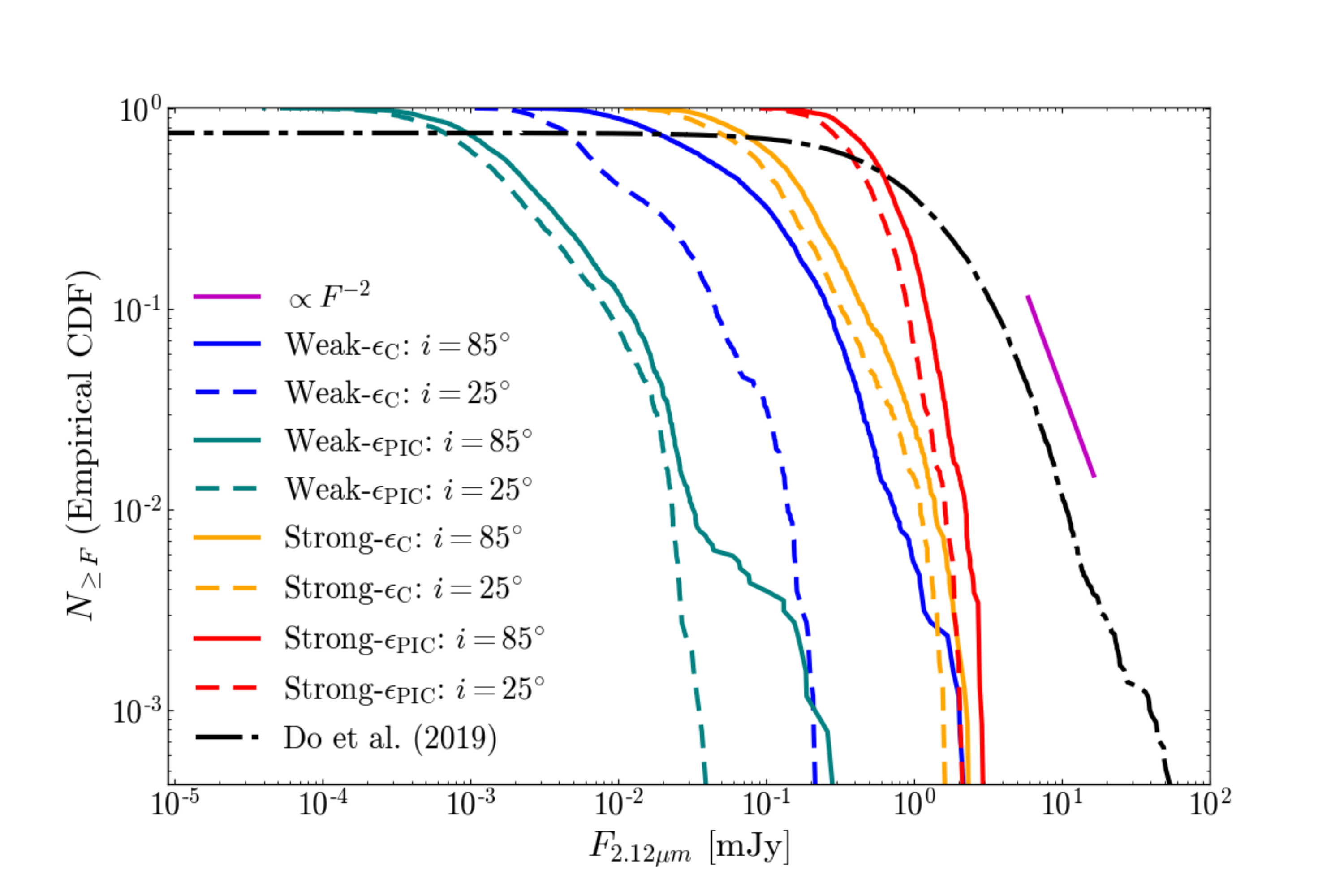}
    
    \caption{Standard GRMHD models are unable to reproduce the NIR flux distribution. We show the $2.12~\mu$m (near-infrared; NIR) flux cumulative distribution functions (CDFs) for each radiative model along with the \citet[][]{Do_2019} measured CDF for \sgra{}. While none of the models provide a satisfactory match to the observed CDF, the strong-field models, \strongC{} and \strongPIC{}, display similar low-to-high flux transition but not enough high flux events. Both weak-field models fail to produce sufficient quiescent emission compared to the data, but display a similar shape for the power-law tail. This pronounced power-law tail suggests a highly variable NIR flux, which is also reflected from the high NIR rms\% values in Table~\ref{tab:rms_fits}. Further, the power-law tail is absent at the lower inclination \weak{} models. We chose the y-axis minimum to be 1/2332, which provides us at least 1 snapshot at the highest flux.}
    \label{fig:NIR_CDFs}
\end{centering} 
\end{figure}  

Figure~\ref{fig:lightcurves}(b), \ref{fig:lightcurves_25}(b), \ref{fig:rms}(b) and \ref{fig:NIR_CDFs} show the lightcurves, power spectra and the cumulative distribution functions at 2.12~$\mu$m (NIR) wavelength. We see that the \weakC{} and \strongPIC{} NIR lightcurves display, on average, higher fluxes than that for \weakPIC{} and \strongC{}, repectively. The PSDs for the strong-field models are similar and display a power-law slope close to $\nu_{\rm F}^{-2}$. The weak-field model PSDs are strikingly different from the strong-field cases, exhibiting flatter slopes at high Fourier frequencies. This is consistent with variability in extremely low fluxes being uncorrelated events, in the form of multiple weakly magnetised current sheets in the accretion flow. The $25^{\circ}$ inclination \weak{} model PSDs have slightly steeper slopes at high frequencies than the corresponding high inclination models.

The weak-field NIR lightcurves display high rms variability with rms\% values $\sim 150\%$ for \weakC{} and $300\%$ for \weakPIC{} (Table~\ref{tab:rms_fits}), which stems from the rapidly fluctuating non-thermal component. These values are close to the observed NIR rms amplitude \citep[$\gtrsim 170\%$;][]{Witzel_2018}. This can be seen from Fig.~\ref{fig:lightcurves}(b) as the flux from both models (blue solid line for \weakC{} and teal dash-dotted line for \weakPIC{}) vary by a factor of 100 over the lightcurve duration. In the low inclination case, the mean NIR flux decreases for all models (see Table~\ref{tab:rms_fits}) while the rms\% remains similar. One noticeable change is in the \weakPIC{} rms\% that drops by a factor of 2. 

One way to decrease the rms\% is to account for synchrotron self-Compton (SSC) upscattering of thermal synchrotron photons as this process might contribute significantly during low flux states in the infra-red (e.g., as seen from Fig. 2 in \citealt{Moscibrodzka_2009}, also see \citealt{Eckart:2004}). The importance of low-level flux events in the overall flux distribution is illustrated more clearly in the CDFs, shown in Fig.~\ref{fig:NIR_CDFs}. 

The CDF ($N_{\geq F}$) is defined as the fraction of the total number of GRRT snapshots ($N_{\rm net}$) where the emitted flux at a given frequency exceed or equals a certain threshold flux $F$,

\begin{equation}
    N_{\geq F}=\frac{1}{N_{\rm net}}\sum^{N_{\rm net}}_{i=1} \text{if}(F_{i}\geq F), 
\end{equation}{}

\noindent where $F_{i}$ are the fluxes for each GRRT time snapshot, and $N_{\rm net}=2555$ and $2332$ for the weak-field and strong-field models, respectively. In Fig.~\ref{fig:NIR_CDFs}, we see that the strong-field models display similar quiescent flux levels to those from \citet[][black dashed line]{Do_2019}, but do not show enough high flux events. The lack of high flux events can be explained via the strong-field spectra (Fig.~\ref{fig:spectra}~c and d) where we see that the NIR spectrum shows less variability compared to the weak-field models, and there are not enough high level flux events to skew the CDF towards a power-law. For the weak-field models, the situation is entirely different as there are too few moderate level flux events and hence, the CDF transitions to a steeper power-law at a smaller flux threshold than for the observed CDF. Here, SSC can contribute to the moderate and low flux levels and skew the CDF transition flux threshold to a higher value. Assuming this flare is due to the same processes, more efficient electron acceleration must occur to explain the high flux excursions as seen in \citet[][]{Do_2019}. A log-normal $+$ power-law-tail distribution describes the weak-field model CDFs better than a log-normal distribution, consistent with the results of \citet{Dodds-Eden_2010, Petersen:2020}. The slopes of the individual CDFs are also interesting to note in the context of log-normal versus power-law CDF distributions. All models except \strongPIC{} display a slope close to -2 and also have rms\% exceeding 120\%. The \strongPIC{} CDF, on the other hand, shows a comparatively steeper slope closer to -4, similar to the values found for lightcurves in \citet[][]{Witzel_2012, Witzel_2018}, as well as a relatively low rms of 60\%. It could be that changes in the accretion rate drive the NIR variability in this model, and therefore, give rise to the log-normal flux distribution. 

It is interesting to note that the low inclination \weak{} NIR CDFs do not show the power-law tail. In the case of \weakPIC{}, it could be due to a change in Doppler-boosting of the emission as the background jet sheath flow is predominantly toroidal. For the \weakC{} case, the mean NIR flux drops by an order of magnitude and hence the corresponding CDF changes. The \strong{} model CDFs remain similar to the corresponding high inclination case. This behaviour of the CDFs suggests that the log-normal component of the CDF is independent of the inclination while the power-law tail can change with inclination. Further, from the \weak{} NIR PSDs, it appears that there is a correlation between a shallow PSD slope at high fourier frequencies and a power-law tail in the CDF, i.e., when the PSD slope steepens, the power-law tail disappears. It follows that the varying accretion rate of the turbulent disc material naturally creates low-flux events, while individual strong reconnection events trigger the formation of fast-moving short-lived blobs of plasma that produce the high NIR flux events, in line with the GRAVITY NIR hotspot observations. Semi-analytical work with hotspot models \citep[e.g.,][]{broderick_2006,younsi_2015} has already shown that the lightcurve is strongly inclination-dependent. Thus, higher resolution GRMHD simulations that capture plasmoid formation would be crucial in confirming the inclination dependence of the NIR CDF. 

\subsection{X-rays: 2-8~keV}
\label{sec:Xrays}

\begin{figure*} 
\begin{centering}
    \includegraphics[width=\textwidth,trim={0 0 0 0}, clip]{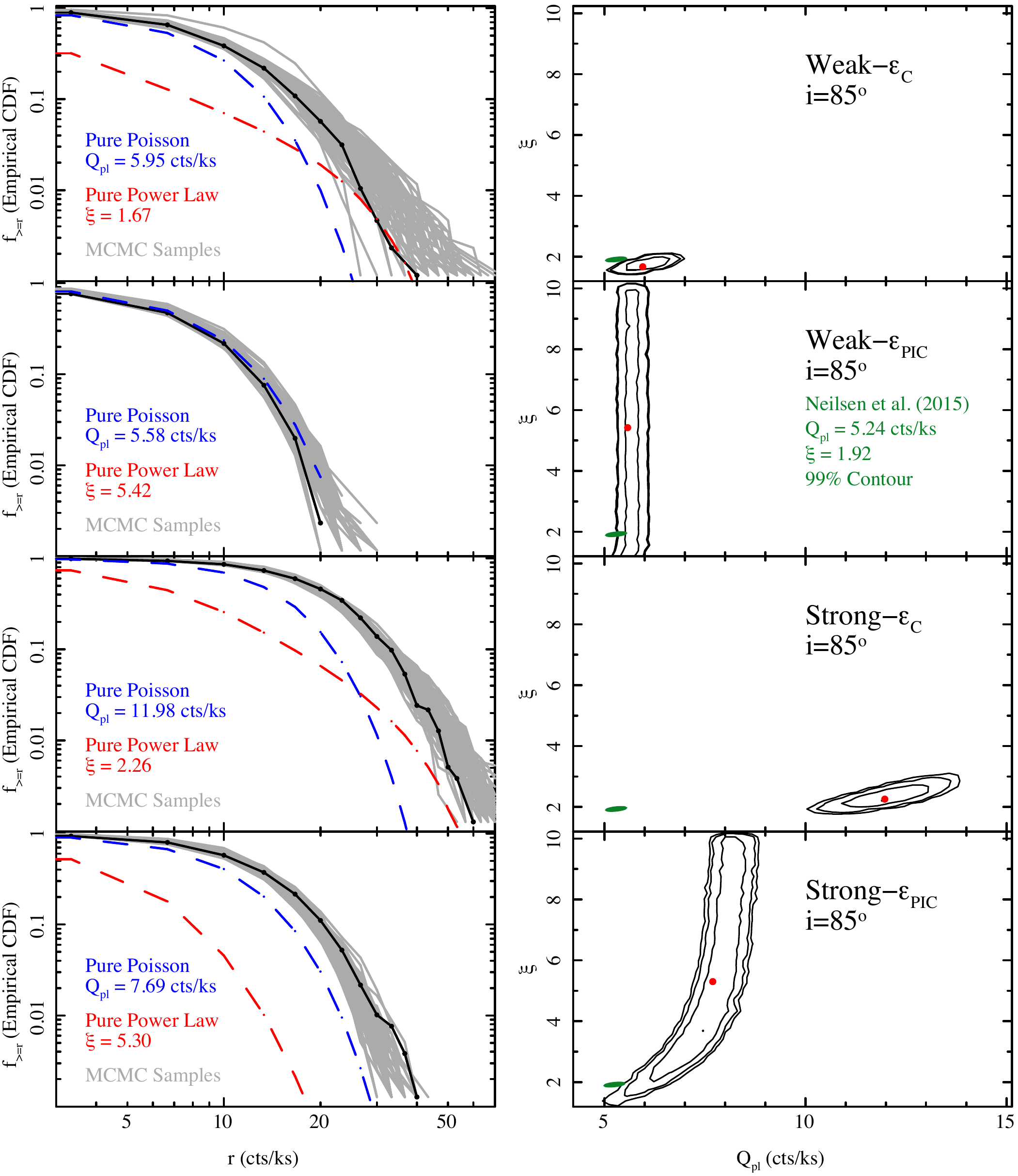}
    
    \caption{We show cumulative distribution functions (f$_{\geq\rm r}$) as a function of the flux count rate (r) for the X-ray 2-8~keV lightcurve constructed from the four different radiative models with inclination angle $85^{\circ}$. The radiative model CDF (black line) is fitted with a Poissonian component (blue-dashed) to represent the quiescent flux distribution and a variable process in the form of a power-law (red-dashed), as indicated by the best-fit values of the Poisson rate Q$_{\rm pl}$ and the power-law index $\xi$ (see Sec.~\ref{sec:Xrays}). In order to directly compare to \sgra{} X-ray CDFs from \citet{Neilsen_2015}, we add a quiescent background count rate to the flux distribution. The grey lines in left panels are the results of MCMC simulations from the joint probability distribution of Q$_{\rm pl}$ and $\xi$. We also show probability contour plots in the right panels that illustrate the correlation between parameters Q$_{\rm pl}$ and $\xi$ (panels b and d), with contours corresponding to 68\%, 90\% and 95\% confidence levels. The green oval indicates the \citet[][]{Neilsen_2015} best fit with the 99\% confidence contour. The \weakC{} parameters fits perform relatively better than the other models when comparing to the \sgra{} best-fit values. All other models either do not show a significant power-law component or have very high Q$_{\rm pl}$. 
    }
    \label{fig:Xray-cdf-85}
    \end{centering}
\end{figure*}
\begin{figure*} 
\begin{centering}
    \includegraphics[width=\textwidth,trim={0 0 0 0}, clip]{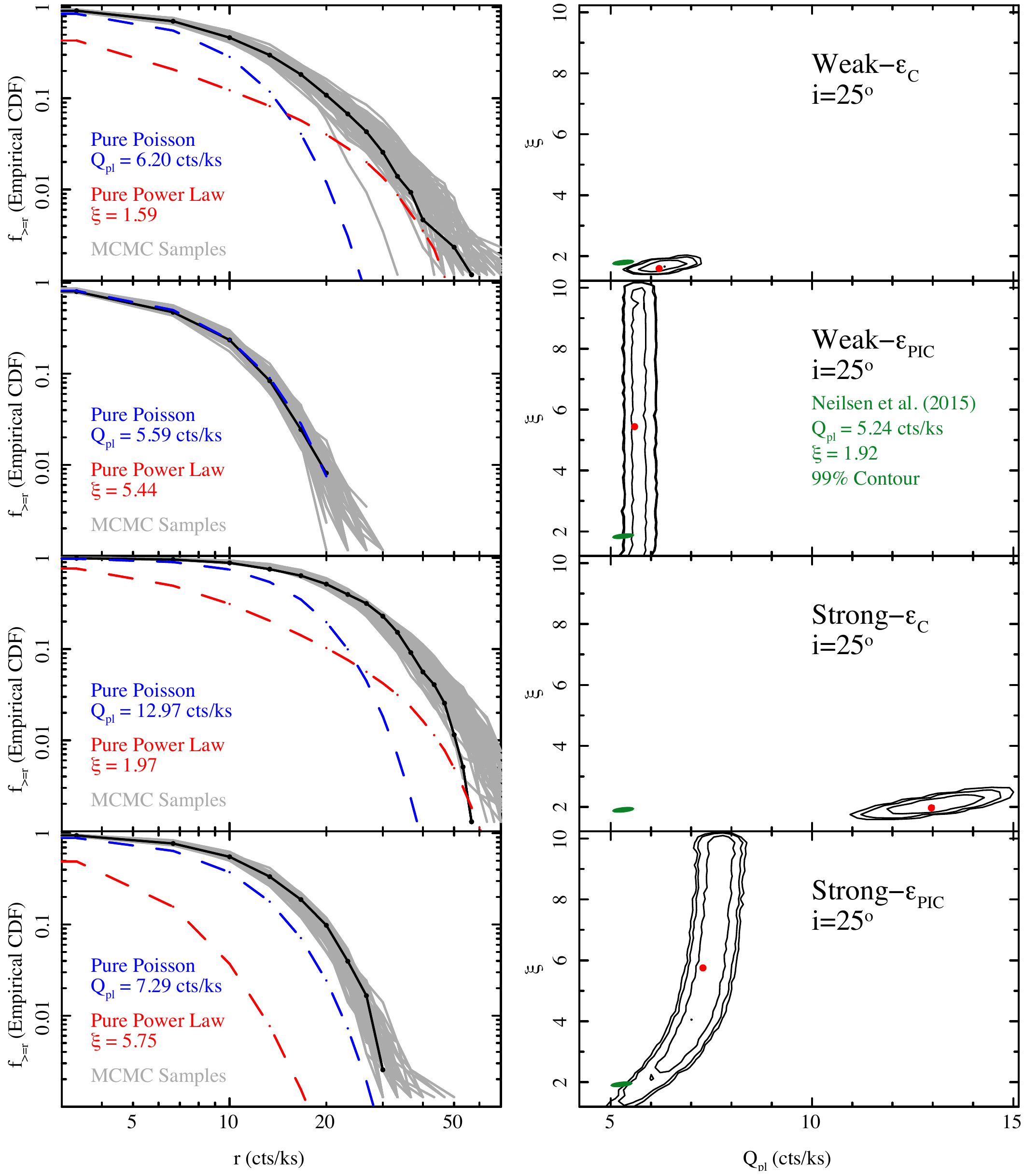}
    \caption{Same as Fig.~\ref{fig:Xray-cdf-85}, but for an inclination angle of $25^{\circ}$. The individual CDFs are similar to their high inclination counterparts.}
    \label{fig:Xray-cdf-25}
    \end{centering}
\end{figure*}

Figures~\ref{fig:lightcurves}(c), \ref{fig:lightcurves_25}(c) and \ref{fig:rms}(c) show the 2-8~keV lightcurves given in units of erg cm$^{-2}$ s$^{-1}$ and the corresponding power spectra for each model, respectively. First we discuss the $85^{\circ}$ inclination lightcurves. From Sec.~\ref{sec:sed}, we see that model \weakC{} displays a mean spectrum over the considered time duration of $\sim 70$~hours that coincides with the total quiescent X-ray emission of \sgra{} from \citet{Nowak_2012}. The lack of any variability in the quiescent spectrum over two decades of observations strongly favours the origin to be thermal bremsstrahlung emission from close to Bondi scales \citep[e.g.,][]{quataert:2002}, which is supported by the resolve extension beyond \chandra{}'s PSF \citep{Wang:2013}. The high mean emission from this model is the outcome of an increase in the accretion rate over the time segment, that also drives a steady increase in the 230~GHz flux (Fig.~\ref{fig:lightcurves}a: blue solid line). On the other hand, \weakPIC{} produces an X-ray lightcurve that is considerably dimmer, even failing to reproduce the expected quiescent non-thermal emission \citep[$\sim 10\%$ of the quiescent emission][]{Neilsen_2013}. The weak-field X-ray lightcurves are less variable as compared to their NIR counterparts, with a rms amplitude of $\sim 144\%$ (for \weakC{}; Table~\ref{tab:rms_fits}) and $\sim 275\%$ (for \weakPIC{}). These values are within an order of magnitude of the observed rms amplitude in X-rays ($\sim 100-1000$) for \sgra{}. The strong-field models exhibit similar lightcurves to each other, with \strongPIC{} displaying higher flux levels than \strongC{} almost throughout the chosen time duration. The low rms\% in the \strong{} models indicate that the variability in the spectra for the strong-field models is quite low, suggesting that the strongly magnetised current sheets in the inner accretion disc, (as seen from the low plasma beta in the region, Fig.~\ref{fig:grmhd} bottom row, middle panel), are structurally stable in time despite the turbulence in the disc. The strong-field model PSDs look similar with comparable rms\%, as the origin of the emission is the same in the two radiative models: the low plasma-$\beta$ jet sheath. The small difference in the variability between \strongC{} and \strongPIC{} is more noticeable in their flux distributions, which we calculate next.

To directly compare our X-ray CDFs to those obtained from the \chandra{} 3~Ms \sgra{} 2012 X-ray Visionary Project, we closely follow the methodology given in \citet{Neilsen_2015}. We process our GRRT X-ray lightcurves using the same tools used to analyse observed \chandra{} lightcurves. To this effect, we use the Interactive Spectral Interpretation System \citep[ISIS;][]{Houck:2000} to fold our model lightcurves with High Energy Transmission Grating Spectrometer (HETGS) \chandra{} \sgra{} responses from \citet{Nowak_2012}. Following Sec.~4.1 in \citet{Nowak_2012}, we include interstellar absorption with the model \texttt{TBnew} \citep{Wilms:2000} with \citet{Verner:1996} cross-sections, assuming the hydrogen column density $N_{\rm H}=14.3\times 10^{22}$ cm$^{-2}$. We then calculate the predicted 2-8~keV count rates. We derive the zeroth order count rates as well as the first order count rates for both HETG grating sets, the medium-energy gratings (MEG) and the high-energy gratings (HEG). We reduce the simulated zeroth order count rates for photon pileup assuming the same scaling as in \citet{Neilsen_2015} since this effect may be as strong as 10-15\% for the highest count rates. We then combine the zeroth and first order count rates to calculate the final intrinsic lightcurve (i.e., this does not include the quiescent emission). For the CDF calculation, we assume a quiescent background X-ray count rate of 5.24~counts/ks to represent the bremsstrahlung contribution from larger scales, adding Poisson noise and interpolating the processed lightcurve onto 300~s bins as done in \citet{Neilsen_2015}.

Figures~\ref{fig:Xray-cdf-85} and \ref{fig:Xray-cdf-25} show the obtained CDFs and the 2D probability contour plots for the weak-field and strong-field models. We represent each radiative model CDF (black lines in the figures) as a combination of a Poisson process (blue dashed) with rate Q$_{\rm pl}$ to characterise the low flux level quiescent emission, and a power-law tail (red dashed) with index $\xi$ for the flare emission. Following \citet{Neilsen_2015}, we then run Markov Chain Monte Carlo simulations (grey lines) to find maximum likelihood fits for Q$_{\rm pl}$ and $\xi$ and compare to the best fit values obtained from \chandra{} observations, given in \citet{Neilsen_2015}: Q$_{\rm pl, bf} = (5.24\pm 0.08)$~counts/ks and $\xi_{\rm bf}=1.92^{+0.03}_{-0.02}$. 

\weakC{} exhibits a higher Poisson rate Q$_{\rm pl}=5.95$~counts/ks and smaller power-law index of $\xi=1.67$ as compared to $Q_{\rm pl, bf}$ and $\xi_{\rm bf}$ (Fig.~\ref{fig:Xray-cdf-85}). Hence, \weakC{} slightly overproduces the quiescent flux-level and reasonably describes the flux distribution for flaring events. Due to the limited duration of our lightcurves, decomposing the CDF into the quiescent Poissonian and power-law tail from the quiescent yields errors larger than those from the \chandra{} XVP data. For \weakPIC{}, the background quiescent flux completely dominates the flux distribution, as is expected from the lightcurve. \strongC{} exceeds the \sgra{} Poisson rate Q$_{\rm pl, bf}$ and displays no traces of a power-law flux distribution at high flux levels. \strongPIC{} exhibits a Poisson rate that is larger than Q$_{\rm pl, bf}$ by a factor of $\sim 2$, with a power-law index $\xi=2.26$ that is steeper than the measured index $\xi_{\rm bf}$, indicating that there is an overabundance of high-level fluxes, and few flaring excursions beyond the calculated quiescent level (e.g., the maximum X-ray flare spectrum in Fig.~\ref{fig:spectra}d). It is possible to reduce the acceleration efficiency for models \weakC{} and \strongC{} in order to achieve the correct Q$_{\rm pl}$ value. However models \weakPIC{} and \strongPIC{} are dependent entirely on plasma-$\beta$ and magnetisation, and do not have any free parameters to adjust the X-ray output. 

Overall, only \weakC{} provides a reasonable description of the X-ray flux distribution in \sgra{}. As mentioned in Sec.~\ref{sec:sed}, the X-ray emission in our models is optically thin synchrotron and, thus, inclination-independent. Consequently, the PSDs and the CDFs do not change significantly with inclination (see Fig.~\ref{fig:Xray-cdf-25}). Combining this information with the absence of the power-law tail in some of the NIR CDFs, it is possible that X-ray-emitting blobs that form near the BH might eventually travel along the jet sheath with relativistic speeds and begin to emit in the NIR band. This is one way that Doppler-boosting might play a pivotal role in determining the CDF shape.

\section{Discussion and conclusions}
\label{sec:conclusions}

\begin{table*}
\begin{center}
\renewcommand{\arraystretch}{1.3}
\begin{tabularx}{1.3\columnwidth}{l | c c c | c c c}
\hline\hline
\vspace*{0mm}
Summary Table  \\
\hline\hline
Model & & NIR & & & X-ray &   \\
\hline
& quiescent flux & CDF slope & rms\% & Q$_{\rm pl}$ & CDF slope $\xi$   & rms\% \\

\hline\hline
Inclination: $85^{\circ}$\\
\hline

\weakC{}  & $\Downarrow$ & Pass & Pass  & Pass  & Pass & Pass\\
\weakPIC{} & $\Downarrow$ & Pass & $\Uparrow$  & Pass   & $\Uparrow$ & Pass\\
\strongC{} & $\Downarrow$ & Pass & Pass  & $\Uparrow$  & Pass & $\Downarrow$\\
\strongPIC{} & Pass & Pass & $\Downarrow$  & $\Uparrow$  & $\Uparrow$ & $\Downarrow$\\

\hline\hline
Inclination: $25^{\circ}$\\
\hline

\weakC{}  & $\Downarrow$ & Pass & Pass  & Pass  & Pass & Pass\\
\weakPIC{} & $\Downarrow$ & Pass &  Pass  & Pass   & $\Uparrow$ & Pass\\
\strongC{} & $\Downarrow$ & Pass & Pass  & $\Uparrow$  & Pass & $\Downarrow$\\
\strongPIC{} & Pass & Pass & $\Downarrow$  & $\Uparrow$  & $\Uparrow$ & $\Downarrow$\\

\hline\hline
\end{tabularx}
\end{center}
\caption{Table qualitatively summarising NIR and X-ray variability results. We distinguish three levels: too low ($\Downarrow$), within reasonable range (``Pass'') and too high ($\Uparrow$). For the reference values, we took NIR quiescent flux $\approx$ 1.1 mJy, NIR CDF slope $\approx$ -2, NIR rms\%$\approx$170\%, X-ray quiescent count rate Q$_{\rm pl}=5.24$~cts/ks, X-ray CDF power-law index $\xi=1.92$ and the X-ray rms\%$\gtrsim 100-1000\%$. Overall, weak-field disc models fare better than strong-field discs. Inclination does not appear to play a significant role.
}
\label{tab:summary}
\end{table*}

The flaring activity in \sgra{} observed in the NIR/X-ray bands provides important clues about the physics of particle energisation and acceleration in the inner few gravitational radii around a SMBH. Current sheets occur naturally in the turbulent, magnetised regions of the disc and will inevitably lead to magnetic reconnection of field lines. Magnetic reconnection results in thermal heating of both electrons and ions, and accelerates a fraction of the electron population to a non-thermal power-law distribution. The nearest SMBH, \sgra{}, is monitored well enough to extract statistical information about the nature of flaring events, e.g., the observed flux distributions and power spectra in the sub-mm, NIR and X-ray wavebands. Here, we take advantage of these measured quantities to test whether the combination of synchrotron emission from non-thermal electrons and disc/jet turbulence can explain the general properties of the flaring events. 

GRMHD simulations are able to capture MHD turbulence in addition to accurately describing the effects of general relativity in the extreme gravity of black holes (Sec.~\ref{sec:grmhd_setup}). Ideal GRMHD, by definition, does not include dissipative processes such as particle acceleration and the effects of radiation, and hence, to directly compare our simulations with observations of \sgra{}, we rely on an additional layer of modelling for the electron properties, together with GRRT radiative transfer as a post-processing step (Sec.~\ref{sec:GRRT_setup}). Using this procedure, we calculate the multiwavelength spectra of our GRMHD$+$GRRT radiative models (Sec.~\ref{sec:sed}), scaling the black hole mass and distance to that of \sgra{} with an accretion rate of a few $\times 10^{-8}\,M_{\odot}$~yr$^{-1}$ that provides a reasonable match to the observed sub-mm flux at two different inclination angles ($25^{\circ}$ and $85^{\circ}$). We only consider synchrotron emission, both from thermal and non-thermal electron populations. Further, to study the variability of our simulation lightcurves, we derive the power spectra and cumulative distribution functions in 3 wavelengths: 230~GHz (sub-mm), $2.12~\mu$m (NIR) and 2-8~keV (X-rays) and compare with their observed counterparts \citep[][(Sec.~\ref{sec:variability})]{Dexter_2014, Neilsen_2015,Witzel_2018}.  

Table~\ref{tab:summary} shows the variability results from this work in a concise way. We find that one of our models, \weakC{}, describes the data reasonably well: (1) the mean spectrum is within observational quiescent limits, (2) the brightest flare X-ray luminosity matches the average X-ray flare spectrum from \citet{Neilsen_2015}, and (3) the X-ray lightcurve is quite variable and the calculated CDF resembles \sgra{}'s X-ray CDF. However, \weakC{}'s brightest flare X-ray luminosity is 25 times smaller than the \citet{Haggard_2019} \chandra{} flare. Further, the X-ray lightcurve does not look similar to observed lightcurves, where flares usually span over longer timescales as compared to the flares obtained from our models. This could be a result of using the ``fast-light'' approximation in our GRRT method, which could become invalid in the vicinity of the black hole, where the photon travel time becomes comparable to the timescale at which the plasma distribution changes \citep[e.g.,][]{Ball_2021}. The NIR CDF follows a log-normal $+$ power-law-like distribution as expected from observations, but has a lower quiescent flux level than that measured for \sgra{}. This model has a relatively weak magnetic flux content in the disc, otherwise known as the ``standard and normal evolution'' (SANE) model, as opposed to near-magnetically arrested strong-field models. The strong-field models, on average, overproduce the quiescent X-ray limits and exhibit a lower level of variability in the X-ray lightcurves as compared to \sgra{}. Thus we favour disc turbulence in SANE models as a reasonable mechanism for explaining \sgra{}'s average flare properties.

None of our models can account for the extremely bright NIR flaring reported in \citet{Do_2019}, though we do achieve moderately high fluxes. There are three possible explanations as to why. It is probable that one needs a pronounced increase in the accretion rate to explain these flares, as suggested by \citet{Do_2019}. If we consider the strong-field model \strongPIC{}, changes in the accretion primarily drive fluctuating flux levels as seen from the variation in the spectrum (grey region in Fig.~\ref{fig:spectra}d) and the brightest X-ray flare spectrum, where the sub-mm and NIR flux both increase by a factor of $\sim 2$ with respect to the mean spectrum. However, turbulence-driven variability is not enough to explain isolated high luminosity events, and we require GRMHD simulations with enough resolution to capture plasmoid formation. Indeed, as \citet{Gutierrez_2020} suggests, a sufficiently strong non-thermal event may be able to explain the \citet{Do_2019} observations. However, the resemblance between our calculated CDFs and the measured CDF from \citet{Do_2019} suggests that processes that frequently occur in the disc and the jet sheath, such as magnetic reconnection in current sheets and variations in the accretion rate, drive most moderate level NIR flux events (also see \citealt{Petersen:2020}). A third possibility is the uncertainty in our assumed electron temperature model, i.e., the turbulent-heating motivated $R_{\rm high}-R_{\rm low}$ electron temperature model from \citet{Howes:2010,Moscibrodzka_2016}. There are alternate electron temperature prescriptions \citep[e.g.,][]{Moscibrodzka_2009,Dexter_2010,Anantua_2020} as well as electron heating models that consider magnetic reconnection and plasma turbulence \citep{Rowan:2017,Werner_2018,Kawazura:2019,Zhdankin:2019}. Indeed, \citet{Dexter_2020} shows that the radio-to-NIR spectrum depends significantly on whether the electrons get heated by turbulence \citep{Howes:2010} or reconnection \citep{Werner_2018}. Similar to our weak-field models, \citet{Dexter_2020} finds high rms\% in the NIR, though the NIR emission originates from only thermal electrons. Interestingly, they also find that heating due to reconnection offers higher variability. Finally, they favour a magnetically arrested disc (MAD) model with reconnection-based electron heating, as weak-field models fail to produce the observed linear polarisation fraction. MAD disc models offer an interesting alternate source of variability: magnetic eruptions \citep[e.g.,][]{tch11} that lead to significant changes in the disc and jet morphology, and contribute to the NIR variability \citep[][]{Dexter:2020:NIR_MAD,Porth:2020:NIR_MAD}. Further, recent simulations of the large scale evolution of the accretion flow starting from the Bondi sphere up to the event horizon by \citet[][]{Ressler:2020:sgra_MAD} also favour a large magnetic flux near the black hole of \sgra{}. However, MAD models are difficult to motivate due to the apparent absence of a strong jet in \sgra{}. In this work, our goal is to examine variability due to turbulence-driven reconnection in relatively-stable discs, and hence, we leave an exploration of MADs for future work. 

Other alternate sources of variability include radiative cooling \citep[e.g.,][]{Fragile:09, Dibi_2012, Yoon:2020}, jet-wind boundary instabilities \citep[e.g.,][]{mck06jf,brom2016,chatterjee2019} and misalignment between the black hole spin vector and the disc angular momentum vector \citep[][]{dexter_2013,White_2020_tiltedimages,Chatterjee_2020}. It is possible that our conclusions may change with the addition of radiative cooling within the GRMHD simulation. Synchrotron and inverse Compton cooling removes internal energy from the gas and changes the dynamics of the turbulence, thereby altering the disc density and temperature profile, even for accretion rates similar to \sgra{} \citep[][]{Yoon:2020}. Hence, if the electron temperature and the acceleration efficiency in the disc current sheets become small enough to satisfy the quiescent NIR and X-ray limits, strong-field models might become a viable option. Indeed, in strong-field models, collisions between the jet and the disc-wind/environment can lead to pinch and kink mode instabilities that efficiently dissipate magnetic energy as heat and/or accelerate particles, as well as lead to enhanced gas entrainment into the jet \citep[e.g.,][]{chatterjee2019}. Finally, misalignment between the black hole spin and the disc may lead to possible turbulent heating events as well as enhanced jet-wind collisions \citep[][]{dexter_2013,White_2020_tiltedimages,Chatterjee_2020}. In order to accurately track particle acceleration due to the reconnection events around black holes, we would require general relativistic PIC simulations \citep[e.g.,][]{Crinquand2020}. Alternatively, in GRMHD, one can use test particles \citep[e.g.,][]{ripperda2019_fluxtube, Bacchini2019} or evolve the electron distribution function \citep[e.g.,][]{Chael:2017, Petersen:2020}. All of these methods improve upon the postprocessing method we use in this work. Advanced models of electron thermodynamics and non-conventional disc geometries will no doubt contribute not only to the overall variability of \sgra{} across its entire multiwavelength emission, but also inform our interpretation of the upcoming 230~GHz Event Horizon Telescope image of \sgra{}. This work is meant to be a first study of non-thermal activity in high resolution 3D GRMHD simulations and further exploration of GRMHD simulations with alternate black hole spins, disc/jet morphologies and orientations, imaged at a variety of inclination angles are required to constrain the possible parameter space of \sgra{} models.

In conclusion, this work presents the first study comparing NIR/X-ray statistics produced from the best available dynamical models of \sgra{} to observations, focusing on variability due to turbulence-driven reconnection. We perform GRRT radiative transfer calculations on two 3D GRMHD models of accreting black holes, one with a weakly magnetised disc and the other a strong-field case, using two different realisations of a hybrid thermal$+$non-thermal electron energy distribution, and generate 230~GHz, $2.12~\mu$m and 2-8~keV lightcurves over a period of time exceeding 60 hours. Table~\ref{tab:summary} shows the time-averaged flux as well as the variability results from our study. A summary of our results is as follows:
\begin{enumerate}
    \item Weakly magnetised discs exhibit high levels of variability in the NIR and the X-rays. Our simulations show that a broken power-law spectrum produces NIR and X-ray CDFs with a slope $\approx -2$, and can be used to explain CDFs of simultaneous NIR/X-ray flares.
    \item Non-thermal synchrotron emission due to disc turbulence in weak-field models explains the average X-ray flare spectrum and flux distribution of \sgra{} reasonably well (see Table~\ref{tab:summary}).
    \item Strongly magnetised discs exhibit low variability as highly magnetised plasma is more abundantly found in the disc and jet sheath. Synchrotron emission from both thermal and non-thermal electron populations contribute to the NIR flux while X-rays originate from non-thermal electrons.
    \item Overall, inclination does not seem to affect NIR and X-ray variability. However, it is possible that Doppler boosting may affect any NIR emission originating in the jet sheath. X-ray emitting plasmoids can travel along the jet sheath, cool and begin to emit in the NIR waveband. 
    \item Disc turbulence alone cannot explain the highly luminous \sgra{} NIR and X-ray flares from \citet[][]{Do_2019} and \citet[][]{Haggard_2019} respectively.
\end{enumerate}
\noindent From our study, it is apparent that we require simulations that are able to resolve the tiny length-scales of plasmoids in order to explain the bright flares seen in \citet[][]{Do_2019} and \citet[][]{Haggard_2019}. We have seen 2D versions of such simulations only recently \citep{Nathanail_2020,Ripperda_2020}. These simulations suggest that we require grid resolutions in excess of 2000 cells across the disc height to trigger the plasmoid instability, indicating that 3D simulations would push us to the brink of computational limitations. Indeed, \citet[][]{Porth:19} showed that variability in the mass accretion rate of weak-field discs decreases with increasing grid resolution, which can have a direct effect on the sub-millimetre variability. It is, however, encouraging to note that our results indicate that one can rely on resolutions similar to this paper to study MHD turbulence-driven variability as a source of moderate level flares in \sgra{}.

\section*{Acknowledgements}
\label{sec:acks}
We thank the anonymous referee for the detailed suggestions that greatly improved the paper. 
This research was enabled by support provided by grant no. NSF PHY-1125915 along with a INCITE program award PHY129, using resources from the Oak Ridge Leadership Computing Facility, Summit, which is a DOE office of Science User Facility supported under contract DE-AC05- 00OR22725, and Calcul Quebec (http://www.calculquebec.ca) and Compute Canada (http://www.computecanada.ca). The authors acknowledge the Texas Advanced Computing Center (TACC) at The University of Texas at Austin for providing HPC and visualisation resources (allocation AST20011) that have contributed to the research results reported within this paper (http://www.tacc.utexas.edu). KC, SM and DY are supported by the Netherlands Organisation for Scientific Research (NWO) VICI grant (no. 639.043.513). K.C. is also supported by a Black Hole Initiative Fellowship at Harvard University, which is funded by grants from the Gordon and Betty Moore Foundation, John Templeton Foundation and the Black Hole PIRE program (NSF grant OISE-1743747). The opinions expressed in this publication are those of the authors and do not necessarily reflect the views of the Moore or Templeton Foundations. ZY is supported by a UKRI Stephen Hawking Fellowship and acknowledges support from a Leverhulme Trust Early Career Fellowship, MvdK is supported by the NWO Spinoza Prize, AT by Northwestern University and by National Science Foundation grants AST-1815304, AST-1911080, and AI by a Royal Society University Research Fellowship. DH acknowledges support from the Natural Sciences and Engineering Research Council of Canada (NSERC) Discovery Grant, the Canada Research Chairs program, and the Canadian Institute for Advanced Research (CIFAR). This research has made use of NASA’s Astrophysics Data System. 

\section*{Data Availability}

Data used to plot the images in this work is available at \href{http://doi.org/10.5281/zenodo.5044837}{http://doi.org/10.5281/zenodo.5044837}.

\bibliographystyle{mnras}
\bibliography{newbib}

\appendix

\section{Non-thermal synchrotron fitting}
\label{sec:Fouka}

\begin{figure*} 
    \includegraphics[width=\columnwidth,trim=0cm 0cm 0cm 0cm,clip]{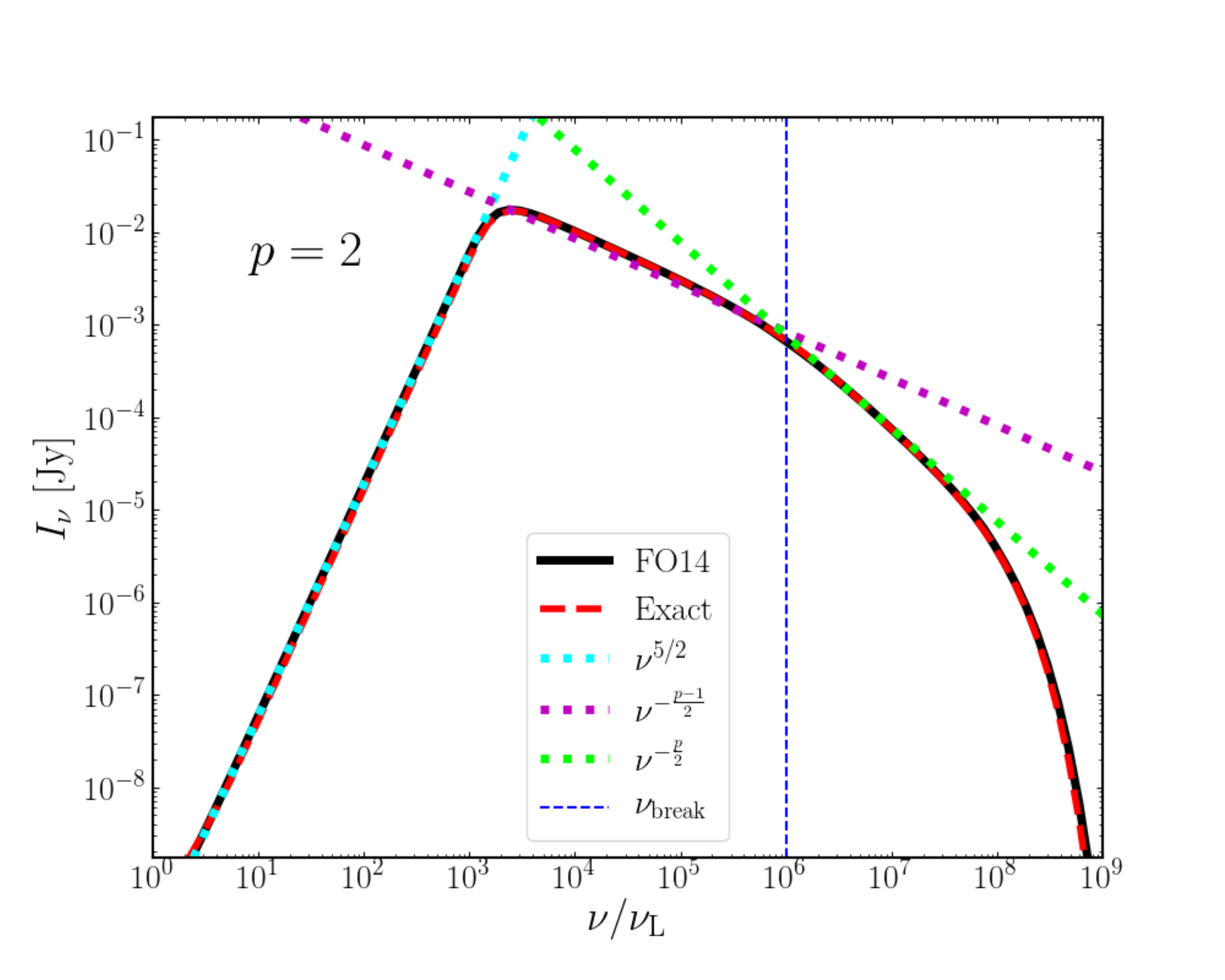}
    \includegraphics[width=\columnwidth,trim=0cm 0cm 0cm 0cm,clip]{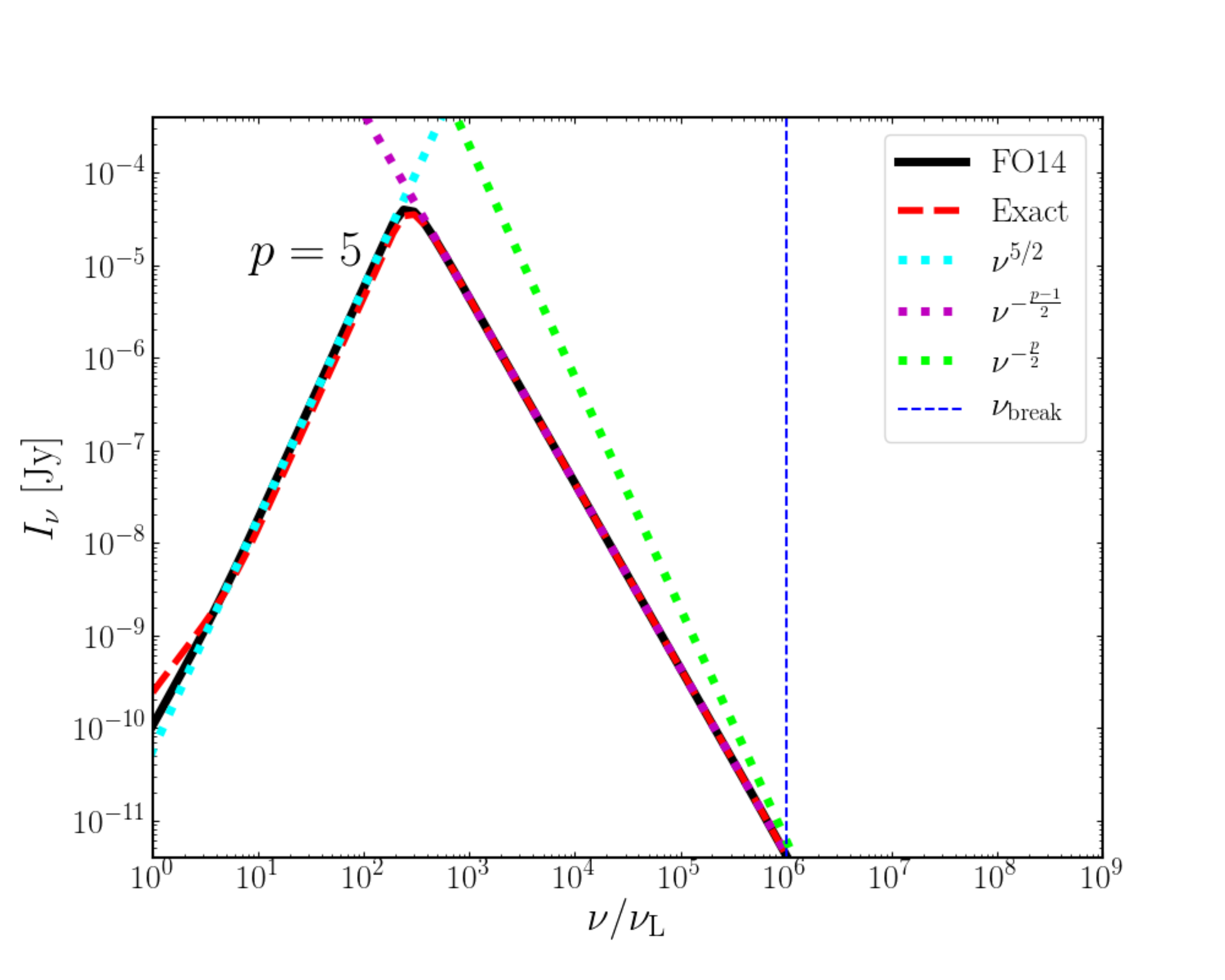}
    \caption{A comparison between the power-law spectrum given by the \citet{Fouka_2014} prescription and the exact solution \citep[e.g.,][]{Rybicki_lightman_book} for a single zone model with power-law indices (left) $p=2$ and (right) $p=5$. See text for the assumed values of the magnetic field, number density and electron Lorentz factor limits, applicable for a current sheet in the accretion flow of Sgr~A$^*$. The \citet{Fouka_2014} prescription matches the exact solution very well within our expected range of power-law indices $1<p<5$.}
    \label{fig:fouka}
\end{figure*}

In this section, we give a description of our synchrotron emissivity and absorption coefficient from a power-law distribution of electrons with power-law index $p$. We take the expressions from \citet{Fouka_2014}. The instantaneous integrated synchrotron spectral power for a pure power-law particle distribution with isotropic pitch angle distribution \citep{Rybicki_lightman_book},
\begin{equation}
    P_{\nu}=\frac{2 \, \pi \, \sqrt{3} \, e^2 \, \nu_{\rm L}}{c}\int^{\gamma_2}_{\gamma_1}d\gamma \, C \, \gamma^{-p} \, F\left(\frac{\nu}{\nu_{\rm c}}\right)
\end{equation}{}
\noindent where $\nu_{\rm L}=eB/(2\pi m_e c)$, $\nu_{\rm c}=(3/2)\gamma^2\nu_{\rm L}$ and $C$ are the Larmor gyration frequency, the synchrotron characteristic frequency, and the normalisation constant respectively. The minimum and maximum limits for the electron Lorentz factor in the power-law distribution are $\gamma_1$ and $\gamma_2$. The synchrotron function $F(\nu/\nu_{\rm c})$ is,
\begin{equation}
    F(z)=z\int^{\infty}_{z}K_{5/3}(\Tilde{z}) \, d\Tilde{z}, \text{ where } z\equiv z(\gamma)=\nu/\nu_{\rm c}.
\end{equation}{}
\noindent The emissivity $j_{\nu}$ and intensity $I_{\nu}$ is given by,
\begin{equation}
    j_{\nu}=\frac{P_{\nu}}{4\pi} \text{ and }
    I_{\nu}=\frac{j_{\nu}}{\alpha_{\nu}}(1-\exp{(-\alpha_{\nu} \, l)})
\end{equation}{}
\noindent where $\alpha_{\nu}$ and $l$ are the absorption coefficient and the photon pathlength respectively. Note that the optical depth is given by $\tau_{\nu}=\alpha_{\nu}l$. \citet{Fouka_2009, Fouka_2014} parameterises the spectral power as $P_{\nu}=P_1F_{\rm p}(x,\eta)$, in terms of a dimensionless frequency $x=\nu/\nu_1$ with $\nu_1=(3/2)\gamma_1^2\nu_{\rm L}$, the Lorentz factor ratio $\eta\equiv \gamma_2/\gamma_1$, and a normalisation coefficient, 
\begin{equation}
    P_1=\pi \, \sqrt{3} \, e^2 \, \nu_{\rm L} \, \gamma_1^{-p+1} \, C/c.
\end{equation}{}
\noindent The parametric function $F_{\rm p}(x,\eta)$ is given as,
\begin{equation}
\begin{centering}
\label{eqn:Fp_x_eta}
F_{\rm p}(x,\eta)= \begin{cases}
    F_{\rm p}(x) \, - \, \eta^{-p+1} \, F_{\rm p}\left(x/\eta^2\right), & \text{for $x<x_{\rm c}$}, \\
    \sqrt{\frac{\pi}{2}} \, \eta^{-p+2} \, x^{-1/2} \, \exp{(-x/\eta^2)}\left[ 1 +  a_{\rm p} \, \frac{\eta^2}{x} \right], & \text{for $x\geq x_{\rm c}$},
    \end{cases}
\end{centering}
\end{equation}
\noindent where $x_{\rm c}=(2.028-1.187p+0.240p^2) \, \eta^2$ and $a_{\rm p}=-0.033-0.104p+0.115p^2$. Here, the \citet{Fouka_2014} fitting formula for $F_{\rm p}$ is given by,
\begin{align*}
    F_{\rm p}\approx & \, \kappa_{\rm p} \, x^{1/3} \, \exp{(a_1 \, x^2 \, + \, a_2 \, x \, + \, a_3 \, x^{2/3})}  \\
    &+ \, C_{\rm p} \, x^{-(p-1)/2}[1 \, - \, \exp{(b_1 \, x^2)}]^{p/5+1/2}
\end{align*}{}
\noindent applicable for $1<p<6$. Here, $\kappa_{\rm p}$ and $C_{\rm p}$, as functions of the Gamma function, are,
\begin{align*}
    \kappa_{\rm p} &= \frac{\pi \,  2^{8/3}}{\sqrt{3} \, (p-1/3) \, \Gamma(1/3)} \\
    C_{\rm p} &=\frac{2^{(p+1)/2}}{p+1} \, \Gamma\left(\frac{p}{4}+\frac{19}{12}\right) \, \Gamma \left(\frac{p}{4}-\frac{1}{12}\right).
\end{align*}
\noindent The coefficients $a_1$, $a_2$, $a_3$ and $b_1$ in terms of $p$ are
\begin{align*}
    a_1 = &-0.14602+3.62307\times 10^{-2}p-5.76507\times 10^{-3}p^2\\
    &+3.46926\times 10^{-4}p^3 \\
    a_2 = &-0.36648+0.18031p-7.30773\times 10^{-2}p^2 \\
    &+1.12484\times 10^{-2}p^3-6.17683\times 10^{-4}p^4  \\
    a_3 = &9.69376\times 10^{-2}-0.48892p+0.14024p^2 \\
    &-1.93678\times 10^{-2}p^3+1.01582\times 10^{-3}p^4  \\
    b_1 = &-0.20250+5.43462\times 10^{-2}p-8.44171\times 10^{-3}p^2 \\
    &+5.21281\times 10^{-4}p^3  
\end{align*}{}

Next, we come to the absorption coefficient, $\alpha_{\nu}=\alpha_1\alpha_{\rm p}(x, \eta)$ for $1<p<5$, where the parametric function $\alpha_{\rm p}(x, \eta)$ and the normalisation coefficient $\alpha_1$ are given as,

\begin{align}
    &\alpha_{\rm p}(x, \eta)  = \, x^{-2} \, F_{p+1}(x, \, \eta) \\
    &\alpha_1  = \, \frac{p+1}{8 \, \pi \,  m_e} \, \nu_1^{-2} \, P_{p+1}(\gamma_1) \, = \, \frac{P_1}{\gamma_1}.
\end{align}

For the cooling case, i.e., the broken power-law electron distribution function, the power-law $p$ changes to $p+1$ at the cooling cutoff, $\gamma=\gamma_{\rm br}$, where $\gamma_{\rm br}$ is calculated by equating the local advection timescale to the synchrotron cooling timescale for electrons (see Eqn.~\eqref{eqn:gamma_br}). Therefore, the distribution function becomes 
 \begin{equation}
\begin{centering}
\frac{d N(\gamma)}{d\gamma}= \begin{cases}
    C_1 \, \gamma^{-p}, & \text{for $\gamma_1<\gamma<\gamma_{\rm br}$}, \\
    C_2 \, \gamma^{-(p+1)}=C_1\gamma_{\rm br}\gamma^{-(p+1)}, & \text{for $\gamma_{\rm br}<\gamma<\gamma_2$}.
    \end{cases}
\end{centering}
\label{eqn:brokenPL}
\end{equation}
\noindent Accordingly, the synchrotron spectral power can be split into two parts,
\begin{equation}
    P_{\nu}=P_1 \, F_{\rm p}\left(\frac{\nu}{\nu_1},\eta_1\right)+P_2 \, F_{p+1}\left(\frac{\nu}{\nu_2},\eta_2\right)
\end{equation}{}
\noindent where,
\begin{equation}
\begin{centering}
\begin{cases}
    \eta_1 &=\gamma_{\rm br} \, / \, \gamma_1,\\
    \eta_2 &=\gamma_2 \, / \, \gamma_{\rm br},\\
    \nu_1 &=\frac{3}{2}\gamma_1^2 \, \nu_{\rm L},\\
    \nu_2 &=\frac{3}{2}\gamma_{\rm br}^2 \, \nu_{\rm L} \, = \, \nu_1 \, \eta_1^2,\\
    P_1 &=\pi\sqrt{3} \, e^2 \, \nu_{\rm L} \, \gamma_1^{-p+1} \, C_1/c,\\
    P_2 &=\pi\sqrt{3} \, e^2 \, \nu_{\rm L} \, \gamma_{\rm br}^{-p} \, C_2/c \, = \, P_1\eta_1^{-p+1}.
\end{cases}
\end{centering}
\end{equation}
\noindent The spectral power can be further simplified as $P_{\nu}=P_1F_{\rm p}(x,\eta_1,\eta_2)$, where $x=\nu/\nu_1$ and the parametric function $F_{\rm p}(x,\eta_1,\eta_2)$ is,
\begin{equation}
    F_{\rm p}(x,\eta_1,\eta_2)=F_{\rm p}(x,\eta_1)+\eta_1^{-p+1}F_{p+1}\left(\frac{x}{\eta_1^2},\eta_2\right).
\end{equation}{}
\noindent The corresponding extension of the absorption coefficient becomes $\alpha_{\nu}=\alpha_1\alpha_{\rm p}(x, \eta_1, \eta_2)$, where,
\begin{equation}
    \alpha_p(x,\eta_1,\eta_2)=\alpha_p(x,\eta_1)+\frac{p+3}{p+2}\eta_1^{-(p+4)}\alpha_{p+1}\left(\frac{x}{\eta_1^2},\eta_2\right).
\end{equation}{}
\noindent Figure~\ref{fig:fouka} shows a comparison between the numerical solution of the radiative transfer equation taken from \citet{Rybicki_lightman_book} and the \citet{Fouka_2014} prescription described above for a single zone model with an electron number density of $5\times 10^7$~cm$^{-3}$, magnetic field strength of 10~G, $\gamma_{\rm min}=\gamma_1=1$, $\eta=\gamma_2/\gamma_1=10^4$ and $\gamma_{\rm br}=10^3$ with power-law indices of $p=2$ and $5$ over a region of $1~r_{\rm g}$ for Sgr~A$^*$. These values are roughly applicable for a current sheet with a size of order of $r_{\rm g}$, and hence, are representative of the conditions around the black hole of Sgr~A$^*$. The non-thermal synchrotron prescription works well within our expect range of power-law index values ($1<p<5$). Beyond $p=5$, we place the acceleration efficiency to be zero as such low power-law indices occur at regions where non-thermal activity is negligible.

\bsp	
\label{lastpage}
\end{document}